\documentclass[12pt,preprint]{aastex}

\usepackage{emulateapj5}
\usepackage{epsfig}

\newcommand{\etal}{\mbox{et al.}}
\newcommand{\ergcms}{erg cm$^{-2}$ s$^{-1}$}
\newcommand{\ergcmsarcmin}{erg cm$^{-2}$ s$^{-1}$ arcmin$^{-2}$}
\newcommand{\ergsec}{erg s$^{-1}$}
\newcommand{\ergsarcmin}{erg s$^{-1}$ arcmin$^{-2}$}
\newcommand{\phcms}{photons cm$^{-2}$ s$^{-1}$}

\newcommand{\msun}{$M_{\odot}$}

\newcommand{\chandra}{{\it Chandra}}

\newcommand{\asca}{{\it ASCA}}

\newcommand{\bepposax}{{\it BeppoSAX}}
\newcommand{\xmm}{{\it XMM-Newton}}

\newcommand{\sgrastar}{\mbox{Sgr A$^*$}}
\newcommand{\program}[1]{{\tt {#1}}}
\newcommand{\html}[1]{{\tt http://#1}}

\newcommand{\gcllb}{\mbox{GRS~1741.9$-$2853}}



\shortauthors{Muno \etal}
\shorttitle{Diffuse X-ray Emission}

\submitted{To appear in ApJ on 20 September 2004}
\begin{document}
\title{Diffuse X-ray Emission in a Deep \chandra\ Image of the Galactic Center}
\author{M. P. Muno,\altaffilmark{1,2} 
F. K. Baganoff,\altaffilmark{3} M. W. Bautz,\altaffilmark{3}
E. D. Feigelson,\altaffilmark{4} G. P. Garmire,\altaffilmark{4}
 M. R. Morris,\altaffilmark{1} S. Park,\altaffilmark{4}
G. R. Ricker,\altaffilmark{3} and L. K. Townsley\altaffilmark{4}}

\altaffiltext{1}{Department of Physics and Astronomy, University of California,
Los Angeles, CA 90095; mmuno@astro.ucla.edu}
\altaffiltext{2}{Hubble Fellow}
\altaffiltext{3}{Center for Space Research,
Massachusetts Institute of Technology, Cambridge, MA 02139}
\altaffiltext{4}{Department of Astronomy and Astrophysics, 
The Pennsylvania State University, University Park, PA 16802}

\begin{abstract}
We examine the spectrum of diffuse emission detected in the 17\arcmin\
by 17\arcmin\ field around \sgrastar\ during 625~ks of \chandra\ observations.
The spectrum exhibits He-like and 
H-like lines from Si, S, Ar, Ca, and Fe, that are consistent with 
originating in a two-temperature plasma, as well as a prominent low-ionization
Fe K-$\alpha$ line. The cooler, $kT \approx 0.8$~keV plasma differs in 
surface brightness 
across the image between $(0.2 - 1.8)\times10^{-13}$~\ergcmsarcmin\ 
(observed, 2--8~keV). This soft plasma is probably heated by supernovae, 
along with 
a small contribution from the winds of massive Wolf-Rayet and O stars.
The radiative cooling rate of the soft plasma
within the inner 20 pc of the Galaxy could be balanced by 1\% of the 
kinetic energy of one supernova every $3\times10^5$ y.
The hotter, $kT \approx 8$ keV component is more spatially uniform, 
with a surface brightness of $(1.5-2.6)\times10^{-13}$~\ergcmsarcmin\ 
(observed; 2--8)~keV. The intensity of the hard plasma is correlated 
with that of the 
soft, but they are probably only indirectly related, because neither 
supernova remnants nor WR/O stars are observed to produce thermal plasma 
hotter than $kT \approx 3$ keV. Moreover, a $kT \approx 8$ keV plasma would
be too hot to be bound to the Galactic center, and therefore would form a 
slow wind or fountain of plasma. The energy required to sustain such a 
freely-expanding plasma within the inner 20 pc of the Galaxy is 
$\sim 10^{40}$~\ergsec. This corresponds to the entire kinetic energy of one 
supernova every 3000 y, which is unreasonably high. However, alternative
explanations for the $kT \approx 8$ keV diffuse emission are equally 
unsatisfying. The hard X-rays are unlikely to result from undetected point 
sources, because no known population of stellar object is numerous enough
to the observed surface brightness. There is also no evidence that non-thermal 
mechanisms for producing the hard emission are operating, as the expected 
shifts in the line energies and ratios from their collisional equilibrium 
values are not observed. We are left to conclude that either there is a 
significant shortcoming in our understanding of the mechanisms that heat the 
interstellar medium, or that a population 
of faint ($< 10^{31}$ erg s$^{-1}$), hard X-ray sources that are a factor
of 10 more numerous than CVs remains to be discovered.
\end{abstract}


\section{Introduction}

Bright, diffuse X-ray and $\gamma$-ray emission has been observed all along 
the Galactic plane, but is particularly bright toward the Galactic center 
\citep[e.g.,][]{wor82,koy86b,yam96,yam97,ski97,vm98}.
The origin of this emission is uncertain. Unlike the cosmic X-ray background, 
the Galactic ridge emission has not yet been resolved into discrete point
sources. On the one hand, \asca\ observations revealed degree-scale 
differences in the surface brightness of the 
diffuse emission that could only be produced by Poisson 
fluctuations in the numbers of 
undetected point sources if they have a luminosity of $\sim 10^{33}$~\ergsec\ 
\citep{yam96}. On the other hand, \chandra\ observations clearly demonstrate 
that there 
are not enough discrete sources with $L_{\rm X} \gtrsim 10^{31}$~\ergsec\ to 
account for more than 10\% the Galactic ridge
X-ray emission \citep{ebi01}. However, the strengths of Fe lines observed 
from the diffuse 
emission are similar to those observed  from Galactic X-ray point sources, 
which suggests that they could be one and the same (Wang, Gotthelf, \& Lang
2002)\nocite{wgl02}. Furthermore,
discrete yet extended sources could produce the spatial variations in
the diffuse emission. Several classes of extended
features have recently been identified, including: regions of bright iron 
fluorescence that are ascribed to molecular clouds being illuminated by 
X-rays from a bright point source \citep{smp93,mur00} 
or bombarded by low-energy cosmic-ray electrons \citep{ylw02};
arcminute-scale features with hard spectra that resemble supernova shocks
\citep{bam03};
and X-ray counterparts to radio features that are thought to be magnetic 
filaments \citep[][Lu, Wang, \& Lang 2003]{sak03}\nocite{lwl03}. 
\chandra\ and \xmm\ observations are only 
beginning to establish how much flux faint point sources and discrete extended 
features contribute to the diffuse Galactic X-ray emission.

If the Galactic ridge X-ray emission is truly diffuse, then it could be 
produced either by hot, $T \ga 10^7$~K plasma or by cosmic-rays interacting 
with neutral material in the interstellar medium (ISM). 
The spectrum of the diffuse emission is one of the most useful 
diagnostics of its origin. \asca\ observations in the 0.5--10~keV band
reveal lines from H-like and He-like ions of Mg, Si, S, and Fe, which indicates
that the diffuse emission cannot originate from a plasma with a single 
temperature \citep{yam96}. As a result, several 
authors have modeled the diffuse emission as originating from two plasma 
components \citep[e.g.,][]{koy96,kan97,tan00}: 
one with $kT_{\rm cool} \approx 0.8$~keV (which we refer to 
here as the ``cool'' or ``soft'' component), and a second with 
$kT_{\rm hot} \approx 8$~keV (which we refer to as ``hot'' or ``hard''). 
The soft plasma is thought to be produced
by supernova shock-waves \citep{kan97}, which are the largest source
of energy for heating the ISM \citep{schl02}.

The origin of the $kT \approx 8$~keV component of the Galactic ridge emission 
is far less certain. The temperature of the putative 
hot plasma is too high for it to be bound to the Galactic disk, so that 
the energy required to sustain it could be equivalent to the release 
of kinetic energy from one supernova occurring every 30~years 
\citep[e.g.,][]{kan97,yam97}. However, supernovae are not observed to produce 
thermal plasma with $kT \gtrsim 3$~keV, and there is no known alternative 
source in the Galactic disk for that much energy. Therefore, several 
alternative sources of the hot plasma have been proposed. 
One possibility is that the $kT \approx 8$ keV plasma is heated by magnetic
reconnection in the ISM, and subsequently confined to the Galactic plane 
by a large-scale toroidal field \citep{tan99}. It is also possible that 
the hot diffuse X-ray emission is a low-energy extension of the emission 
with a power-law spectrum observed above 10~keV, which suggests that it 
may result from a non-thermal mechanism 
\citep[see][but see Lebrun \etal\ 2004]{ski97,yam97,vm98}. 
Among the proposed mechanisms are charge-exchange interactions between 
cosmic ray ions and interstellar matter 
(Tanaka, Miyaji, \& Hasinger 1999; Valinia 2000)\nocite{tmh99,val00},
bremsstrahlung radiation from cosmic-ray electrons or
protons \citep{vm98}, and quasi-thermal emission from plasma that is 
continuously accelerated by supernova shocks propagating in the  
$kT \approx 0.8$~keV component of the ISM \citep{dog02,mas02}. These 
non-thermal processes should produce line emission with energies and flux
ratios that differ significantly from those expected from a plasma in 
thermal equilibrium. Unfortunately, previous \asca\ observations were unable to
determine the ionization state of the plasma unambiguously, because the 
spectrum was contaminated by an instrumental Fe line between 6--7~keV.

Clearly, further study of the diffuse
X-ray emission is important for understanding stellar life cycles, 
magnetic fields, and cosmic ray production in the Galaxy.
In this paper, we study the spectral properties of the diffuse X-ray 
emission from a 17\arcmin\ by 17\arcmin\ field around \sgrastar\ that
has been observed for over 600~ks with \chandra. 
These observations
have several advantages over previous ones with \asca\ \citep{koy96} 
and \bepposax\ \citep{sm99}: 
(1) the long integration provides sufficiently large signal-to-noise 
ratio to study 
the spectrum of the diffuse emission from arcminute-scale sub-regions of 
the field, (2) the 0\farcs5 
angular resolution allows us to resolve the truly diffuse emission from 
filamentary features and point sources, and (3) the relative lack of 
instrumental lines, particularly between 6--7~keV, allows us to measure the 
ionization state of the diffuse emission with greater confidence.
The layout of the paper is as follows. In Section~2.1 we present images
that provide an overview of the diffuse emission from 
the field. In Section~2.2, we examine how the spectrum of the diffuse emission 
differs across the field. In Section~2.3, we compare the spectra of the
point sources and diffuse emission, and place upper limits on the contribution
of undetected point sources to the diffuse emission. In Section~3.1, we
derive the properties of the putative plasma responsible for the diffuse 
emission. These are used in Sections~3.2 and 3.3 to examine the likely origins
of the diffuse emission. In Section~3.4, we discuss the number of undetected
point sources that may be present in the field. Finally, in Section~4, we 
list the contributions of point sources, extended features, and diffuse 
emission to the X-ray luminosity of the Galactic center. 

\section{Observations}

The inner 20 pc of the Galaxy have been observed on twelve occasions 
as of June 2002 (Table~\ref{tab:obs}) using the Advanced CCD 
Imaging Spectrometer imaging array (ACIS-I) aboard 
the {\it Chandra X-ray Observatory} \citep{wei02}. 
The ACIS-I is a set of four 1024-by-1024 pixel CCDs, covering
a field of view of 17\arcmin\ by 17\arcmin. When placed on-axis at the focal
plane of the grazing-incidence X-ray mirrors, the imaging resolution 
is determined primarily by the pixel size of the CCDs, 0\farcs492. 
The CCDs also 
measure the energies of incident photons, with a resolution of 50--300 eV 
(depending on photon energy and distance from the read-out node) within a 
calibrated energy band of 0.5--8~keV.

We reduced the 
data from each observation using the methods described in detail in 
\citet{mun03} and \citet{tow03}. In brief, we first created 
an event list for each observation in which we corrected the pulse heights of 
each event for the position-dependent charge-transfer inefficiency 
\citep{tow02a}. Next, we excluded events that did not pass the 
standard ASCA grade filters and CXC good-time filters. We then removed 
intervals in which the count rate from the diffuse emission flared 
to 3-$\sigma$ above the mean rate in the 0.5--8.0 keV band, presumably due 
to particles impacting the detector. This removed all excursions larger than 
$\approx 15$\%, which represented a total time of only 15~ks (2\% of the
total exposure). The remaining long-term variations in the particle 
background should be no larger than 5\% \citep[see][]{mar03}. 
The final total live time was 626 ks. 

\begin{figure*}[th]
\centerline{\epsfig{file=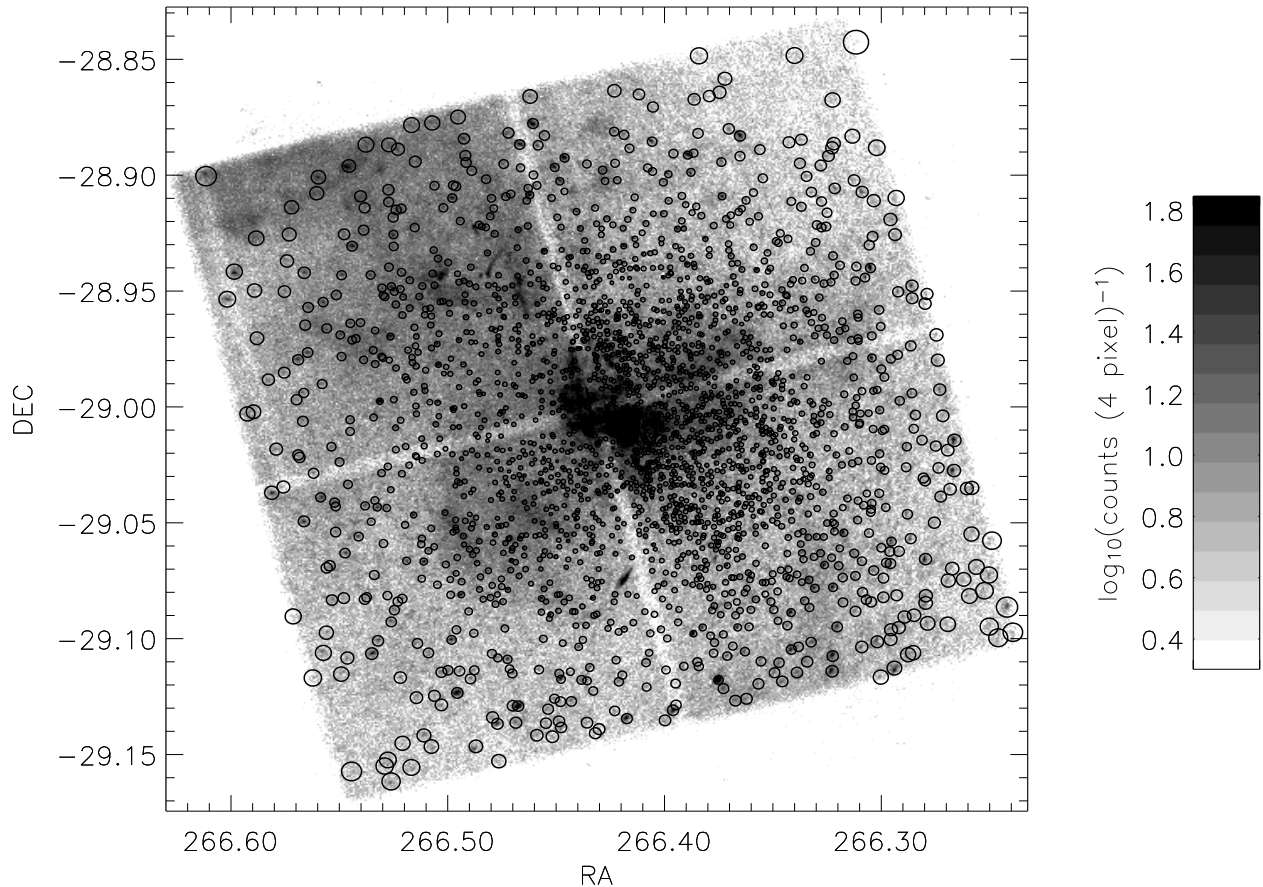,width=0.95\linewidth}}
\caption{
Raw image of counts received in the \sgrastar\ field. The image has been binned
to have 2\arcsec\ pixels. We have overlaid on the image the circles that were 
used to exclude the 2353 point sources that we detected \citep{mun03}.}
\label{fig:rawimg}
\end{figure*}

Before extracting spectra of the diffuse emission, we needed to remove the
point sources from the image.
To identify the point sources, we created a composite event list
by re-projecting the sky coordinates 
of the events from each observation onto the plane tangent to the radio 
position of \sgrastar. We excluded the first half of ObsID 1561 from the
composite event list because a bright $10^{-10}$~\ergcms\ transient 
dominated the northwest portion of the field 
\citep[\gcllb; see][]{mun03b}. An image based upon this event list 
 is displayed in Figure~\ref{fig:rawimg}. We searched for point sources
in images of events in three energy bands (0.5--8~keV, 
0.5--1.5~keV, and 4--8~keV) using \program{wavdetect}. We used a 
significance threshold of 
$10^{-7}$, which corresponds to the chance of spuriously identifying 
Poisson fluctuations within a beam defined by the
instrumental point spread
function (PSF).\footnote{We 
note that this significance was designed to limit the 
number of spurious sources, and therefore some weak point sources that are 
difficult to distinguish from statistical fluctuations in the background 
diffuse emission are not included in the catalog. We ran searches with less 
strict thresholds, and only identified a couple hundred additional sources. 
These sources contribute less than 0.5\% to the total counts in the diffuse 
emission, so we let them remain in the image.}
We identified a total of 2357 X-ray point sources 
(see Table~3 in Muno et al. 2003a), of which 1792 were detected 
in the full band, 281 in the soft band (124 exclusively in the soft band), 
and 1832 in the hard band (441 exclusively in the hard band). 

To isolate the diffuse emission in each observation, we excluded events that 
fell within circles circumscribing the 95\% contour of the PSF around each 
point source. 
The excluded regions are indicated with circles in Figure~\ref{fig:rawimg},
and range in size from 3\arcsec\ within 4\arcmin\ of the aim-point to 
7\arcsec\ at an offset of 7.5\arcmin\ from the aim-point.
We found that the 95\% contour struck an appropriate balance 
between removing most of the counts from point sources in the image, and 
leaving diffuse emission to analyze. For instance, in the inner few
arcminutes of the image, the density of point sources is so high that the 
circles circumscribing the 99\% contour of the PSF cover the image, leaving 
few counts that are unambiguously diffuse emission. Few photons from the 
observed point sources should contaminate the diffuse emission. There
are only $2\times10^5$ net counts from the point sources in the catalog of 
\citet{mun03}, compared to $2\times10^{6}$ counts in the diffuse
emission, so fewer than 0.5\% of the counts in the diffuse emission should come
from the wings of the PSFs for the detected point sources. The flux contributed
by the dust-scattering halos of the observed point sources can be estimated 
using the optical depth toward sources near \sgrastar\ from 
\citet[][their Figures 1 and 8]{td04}, and the expected energy-dependent 
profiles of the halos from \citet[][their Figures 9 and 10]{dra03}. We 
estimate that at 7\arcmin\ from the aim-point the scattered 
component equals the observed point-source flux at 2~keV, but declines to
$\la 20$\% of the point-source flux above 4~keV. Within 4\arcmin\ of the
aim-point the exclusion regions for the point sources are smaller, so the 
contribution from scattered flux is twice as large. However, point
sources produce only 4\% of the diffuse flux at 2~keV and 15\% above 4 keV
(see Section~\ref{sec:ps}), so the scattered flux contributes only a few 
percent to the diffuse emission. For comparison, the systematic 
uncertainty in the calibration of the CTI-corrected ACIS response is similar, 
on order 3\%. 

Finally, we removed the bright, filamentary features identified 
in the Northeast portion of the image by \citet{par03}. These features 
contributed $\approx 15$\% of the 
\begin{figure*}[p]
\centerline{\epsfig{file=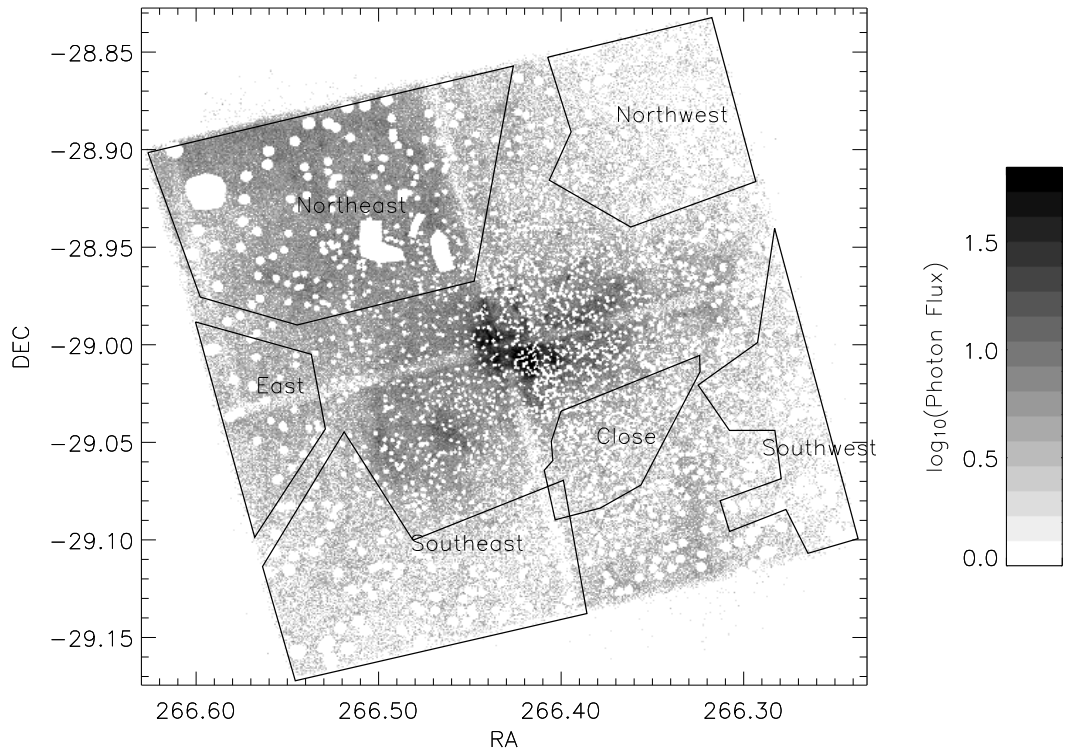,width=0.8\linewidth}}
\caption{
Image of the \sgrastar\ field in the 2--4~keV band, overlaid with the 
polygonal regions from which 
the spectra of the diffuse emission were extracted. The ``holes'' apparent
in the data result from removing events near point sources and obvious 
filamentary features.}
\label{fig:softimg}
\end{figure*}
\begin{figure*}[p]
\centerline{\epsfig{file=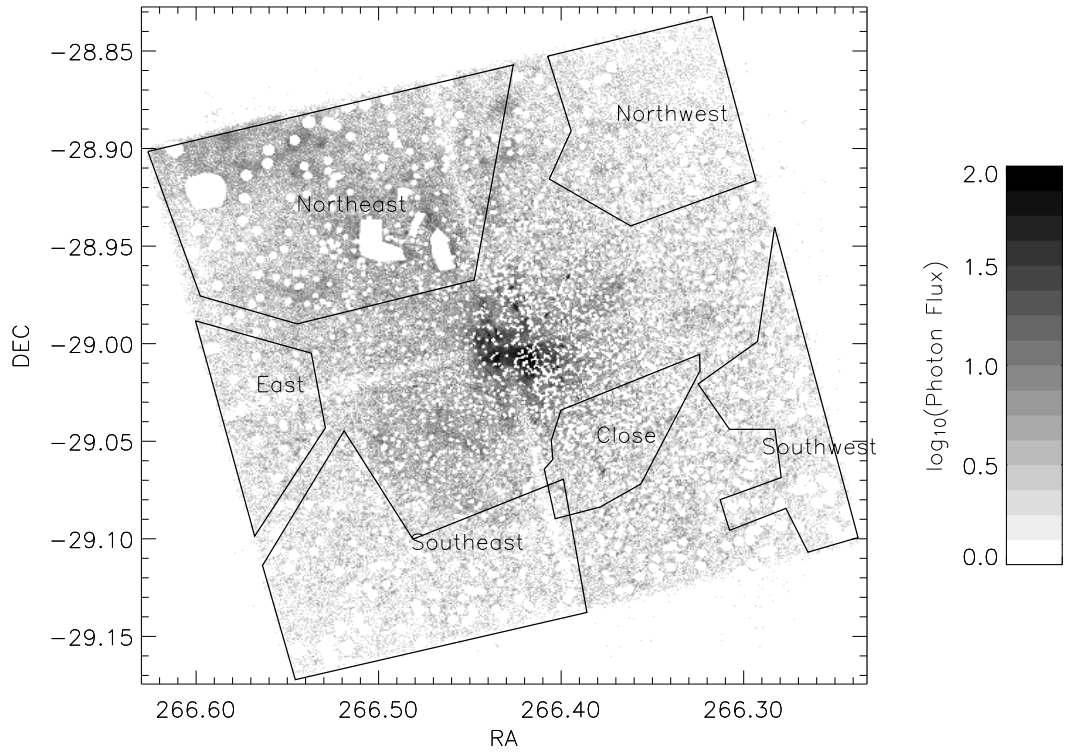,width=0.8\linewidth}}
\caption{
Same as Figure~2, for the 4--8~keV band.}
\label{fig:hardimg}
\end{figure*}
flux from this region. We used the 
resulting event lists to create images and spectra of the diffuse emission.

\begin{figure*}[th]
\centerline{\epsfig{file=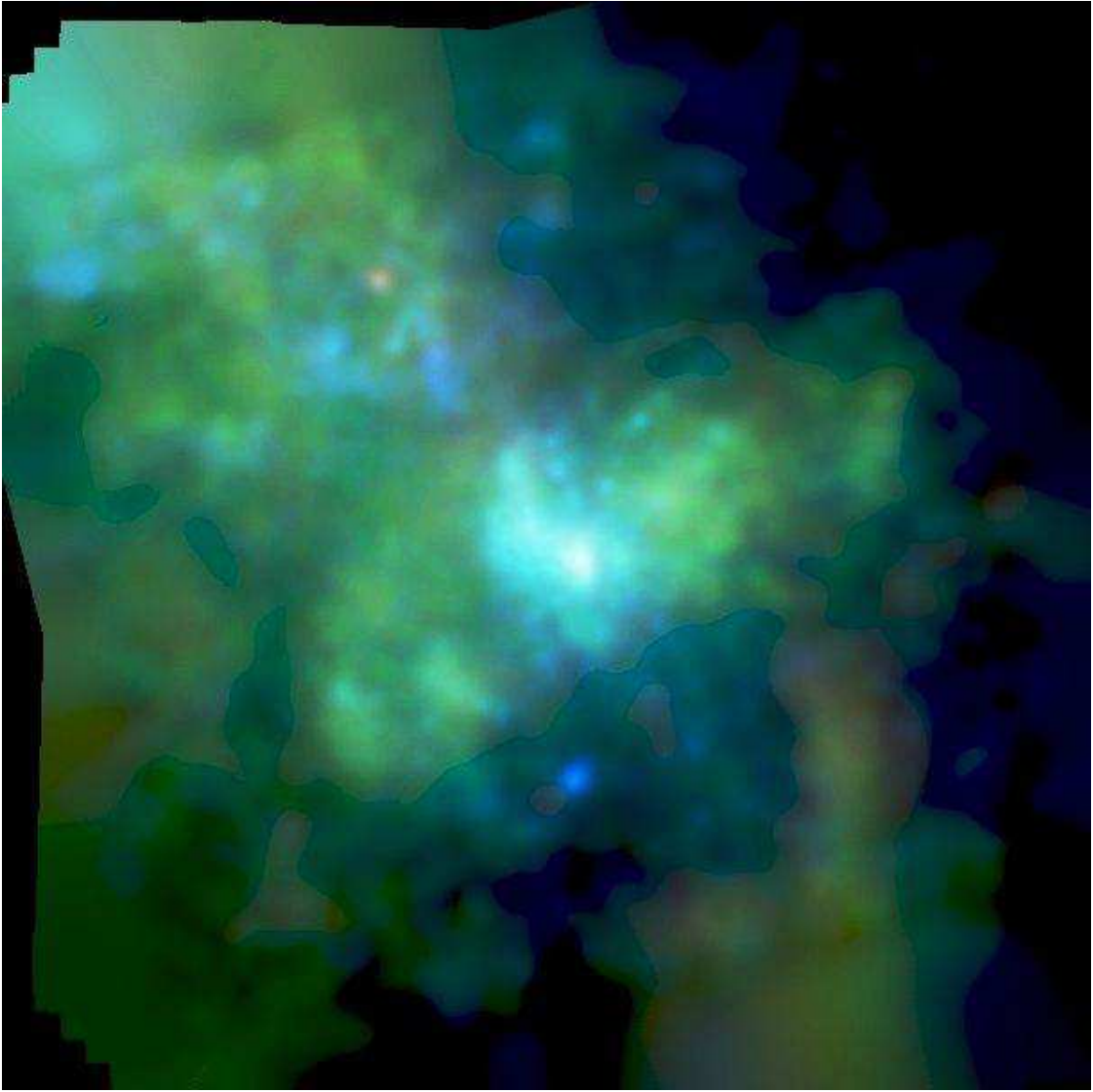,width=0.8\linewidth}}
\caption{
Three-color, smoothed image of the \sgrastar\ field. The red band 
is made from photons between 0.5--2.0~keV, the green from 2.0--4.0~keV, 
and the blue from 4.0--8.0~keV. Point sources were removed from the 
image as in Figures~2 and 3, and the image was adaptively smoothed using
the techniques of \citet{tow03}. The intensities of each band were 
scaled to highlight faint, diffuse features.}
\label{fig:twocolor}
\end{figure*}

\subsection{Images of the Diffuse Emission}

We combined the event lists from each observation to produce the images
of the diffuse emission in the 2--4~keV and
4--8~keV bands in Figures~\ref{fig:softimg} and
\ref{fig:hardimg}. The ``holes'' apparent in the images were left by removing
the events associated with the point sources and filamentary features. 
These images provide a qualitative understanding of the diffuse
emission.

In the soft band (2--4~keV), the inner part of the image is dominated by
\sgrastar, the nuclear stellar cluster, and
Sgr A East. Beyond these, two 
lobes of X-ray emission are oriented perpendicular to the
Galactic plane and centered on \sgrastar. It is likely that these represent
an outflow from the central parsec \citep{bag03,mor03}. 
Enhanced X-ray emission is 
also evident in the Northeast portion of the image, between the Galactic center
and the Radio Arches region located $\approx 3$ arcmin to the north
of the image \citep{lar00,wgl02}. This emission exhibits
prominent He-like lines from Si and S, and low-ionization K-$\alpha$ 
emission from Fe \citep{par03}.  
Finally, a broad ridge of emission with low surface brightness is evident 
to the southwest (lower-right). 

The hard band (4--8~keV) is also dominated by the Sgr A complex at the center
of the image. At larger radii from the center, the most prominent features 
are filaments dominated by low-ionization
Fe emission at 6.4~keV in the Northeast \citep{par03}, and a hard, 
continuum-dominated feature in the south \citep{sak03,lwl03,yz03}. These 
features have been removed from the image, as they were not included when
we modeled the diffuse emission. Enhancements
in the hard diffuse emission are also observed at the bipolar lobes, and in
the Northeast. 

In Figure~\ref{fig:twocolor}, we display a smoothed, three color image of 
the field. The red band was made using photons between 0.5--2.0~keV, 
the green with 2.0--4.0~keV, and the blue
\begin{figure*}[th]
\centerline{\epsfig{file=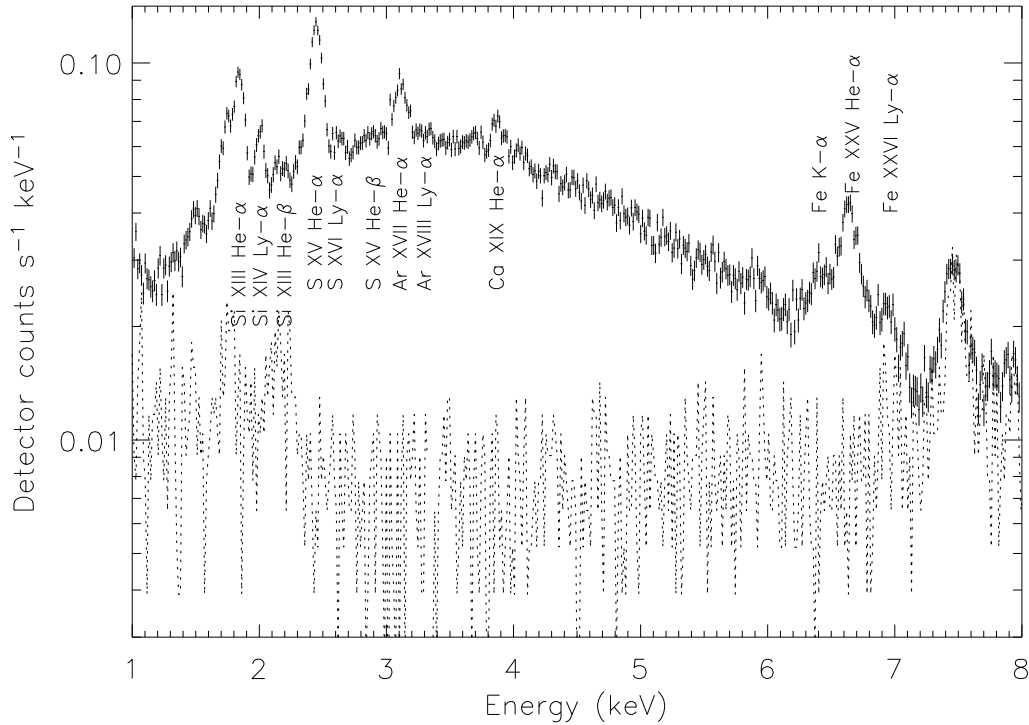,width=0.8\linewidth}}
\caption{
Spectrum of the diffuse emission from the Southeast dark region, before 
background subtraction. The instrumental background is indicated 
with the grey dashed line. Prominent lines from the diffuse emission
are indicated.}
\label{fig:diffraw}
\end{figure*}
 with 0.4--8.0~keV. After removing 
the point sources, the image was adaptively smoothed using the algorithm 
described by 
\citet{tow03}. The red band contributes very little flux to the image, because
the Galactic absorption column prevents us from receiving photons with 
energies $\lesssim 2$~keV from the Galactic center. However, soft photons
are received from the Sgr A complex, and from a bright, slightly extended 
feature of uncertain nature about 6\arcmin\ north-northeast of \sgrastar. 
All of the other features evident in the un-smoothed images are apparent in 
the smoothed image.

\subsection{Spectra of the Diffuse Emission\label{sec:spec}}

We used the images in Figures~\ref{fig:rawimg}--\ref{fig:twocolor} to 
select regions from which we extracted spectra of the diffuse emission. The
regions are displayed with polygons in Figures~\ref{fig:softimg} and
\ref{fig:hardimg}. Five of the regions were chosen because they were 
particularly dark. The sixth was chosen from the bright region in 
the Northeast to help understand the nature of the surface brightness 
variations in Figure~\ref{fig:twocolor}. We avoided the bipolar lobes
and the ridge to the southwest, as they will be studied elsewhere.

In order to model the spectrum, for each region and each observation
we computed an effective area function using the CIAO tool \program{mkwarf}, 
which accounts for the vignetting and satellite dither using the distribution 
of counts received in the extraction region. 
We then computed a weighted average of the effective area for each region, 
using 
as the weighting the number of counts in that region in each observation. 
Similarly, we computed a mean response by averaging the 
response functions provided by \citet{tow02b} from the range of detector
rows covered by each region, also weighted by the
number of counts in 
each observation. We estimated the background produced by particles impacting
the detector using a 50~ks observation taken with the 
ACIS-I stowed out of the focal plane of the mirror 
assembly.\footnote{\html{cxc.harvard.edu/contrib/maxim/stowed/}} 
In order to account for spatial variations in the background across the 
ACIS-I chips, we extracted the 
background events from regions identical to those we used for the spectra of 
the diffuse emission. The background observations had the same CTI correction
and filtering applied as we used for the source events. 
The area, average offset from the nominal aim point (\sgrastar), total counts,
and estimated instrumental background for each region are listed in 
Table~\ref{tab:regions}. The instrumental background represents 20--25\% 
of the total counts in the dark regions, but only 10\% of the total counts 
in the bright region to the northwest. We note that the unresolved 
cosmic X-ray background contributes insignificantly to the diffuse emission. 
If we account for $10^{23}$ cm$^{-2}$ of absorption through 
the Galaxy, and the fact that sources brighter than 
$6\times10^{-15}$~\ergcms\ (2--8~keV; de-absorbed to match the deep-field 
surveys) should have been detected, less than 1\% of the observed diffuse
emission is from extra-galactic background \citep[e.g.,][]{ros02}.

Figure~\ref{fig:diffraw} displays the source and background spectrum
of the Southeast region, which has the highest signal-to-noise of the
dark regions. Many lines are detected in the spectrum: the He-like 
$n=2-1$ transitions of Si, S, Ar, Ca, and Fe; the He-like $n=3-1$ 
transitions of Si and S; 
the H-like $n=2-1$ transitions of Si, S, Ar, and Fe;
low-ionization (``neutral'') Fe K-$\alpha$ at 6.4~keV; and
an instrumental Ni line at 7.5~keV. Background-subtracted spectra are 
displayed in Figure~\ref{fig:diffmod}; the absence of the instrumental Ni line 
at 7.5~keV indicates that the background subtraction was successful, whereas
the other lines are clearly intrinsic to the diffuse emission. 

Motivated by past investigations, we modeled the spectrum of the diffuse 
emission in two ways. First, we modeled the spectrum between 1--8~keV using
two thermal plasma components. Second, we analyzed the spectrum between 
4.5--8.0~keV in order to examine the properties of the iron lines. 

\subsubsection{Two-$kT$ plasma model\label{sec:twokt}}

As pointed out by several authors \citep[e.g.][]{koy86a,yam96,tan00}, 
the prominent line emission in the spectrum suggests that it originates from 
optically thin plasma, whereas the simultaneous presence 
of Fe with Si, S, and Ar
suggests that the plasma has multiple temperature components.
The plasma model we used
(\program{apec}\footnote{\html{cxc.harvard.edu/atomdb}} 
in \program{XSPEC}; see also, e.g., Raymond \& Smith 1977;
Mewe, Lemen, 
\& van den Oord 1986; Borkowski, Sarazin, \& Blondin 1994) assumes that it
is in collisional ionization equilibrium, and is 
able to self-consistently account for both the continuum emission and the lines
from H-like and He-like species. The neutral Fe emission must be 
included ad-hoc as an additional Gaussian component.  

Several assumptions were required for the model to reproduce the data. First,
we only applied the model between 1.0 and 8.0~keV. Below this energy
range the photon counts are dominated by foreground emission, while above 
this range the spectra are background-dominated. Second, we examined
data taken in 2002 May from the on-board calibration sources (which produce 
lines at the K-$\alpha$ and K-$\beta$ transitions from Al, Ti, and Mn when the
detector is exposed to them), and found that even after applying the CTI 
correction the observed line centroids were slightly below their expected 
values. The shift ranged from $\la 0.2$\% near the detector read-out 
(at the top and bottom of Figure~\ref{fig:rawimg}), up to $\approx 0.9$\% 
at the top of each CCD (the center of the image). Therefore, we allowed for a 
$\approx 0.5$\% shift in the energy scale in each
spectrum. When fitting a Gaussian to the 6.4~keV
line of Fe, the line centroids were allowed to vary once, and then 
were frozen.
When using plasma models, the red-shift parameter was used to change the 
energy scale. Third, a 
3\% systematic uncertainty was added in quadrature to the statistical 
uncertainty on the count rate in order to account for 
uncertainties in the ACIS effective area.
Finally, we found that we needed to allow the abundances of Si, S, Ar, Ca, 
and Fe to vary independently to obtain an adequate fit. 

To account for absorption we assumed (1) that the entire 
region was affected by one column of material that represents the average
Galactic absorption (modeled with 
\program{phabs} in \program{XSPEC}) and (2) that a fraction of each region was 
affected by a second column that represents absorbing material with a smaller
filling factor along the line of sight (modeled with \program{pcfabs}). 
The absorption model roughly accounts for the fact that both the plasma and 
absorbing material are distributed along the line of sight
(see, e.g., Vollmer, Zylka, \& Duschl 2003)\nocite{vzd03}.
The mathematical form of the model was
\begin{equation}
e^{-\sigma(E)N_{\rm H}}([1-f] + fe^{-\sigma(E)N_{\rm pc,H}}),
\label{eq:abs}
\end{equation}
where $\sigma(E)$ is the energy-dependent absorption cross-section, 
$N_{\rm H}$ is the absorption column, $N_{\rm pc,H}$ is the partial-covering
column, and $f$ is the partial-covering fraction. 
The absorption was modeled using a separate factor 
(i.e., Equation~\ref{eq:abs}) 
for each plasma component, although, aside from the quality of the 
fits, the basic results do not change if 
we use a single absorption factor for both plasma components.
We note that for the 
soft plasma component the best-fit value of $f$ approached 1 and was 
poorly constrained, so we fixed its value to $f = 0.95$. 

\begin{figure*}[th]
\centerline{\epsfig{file=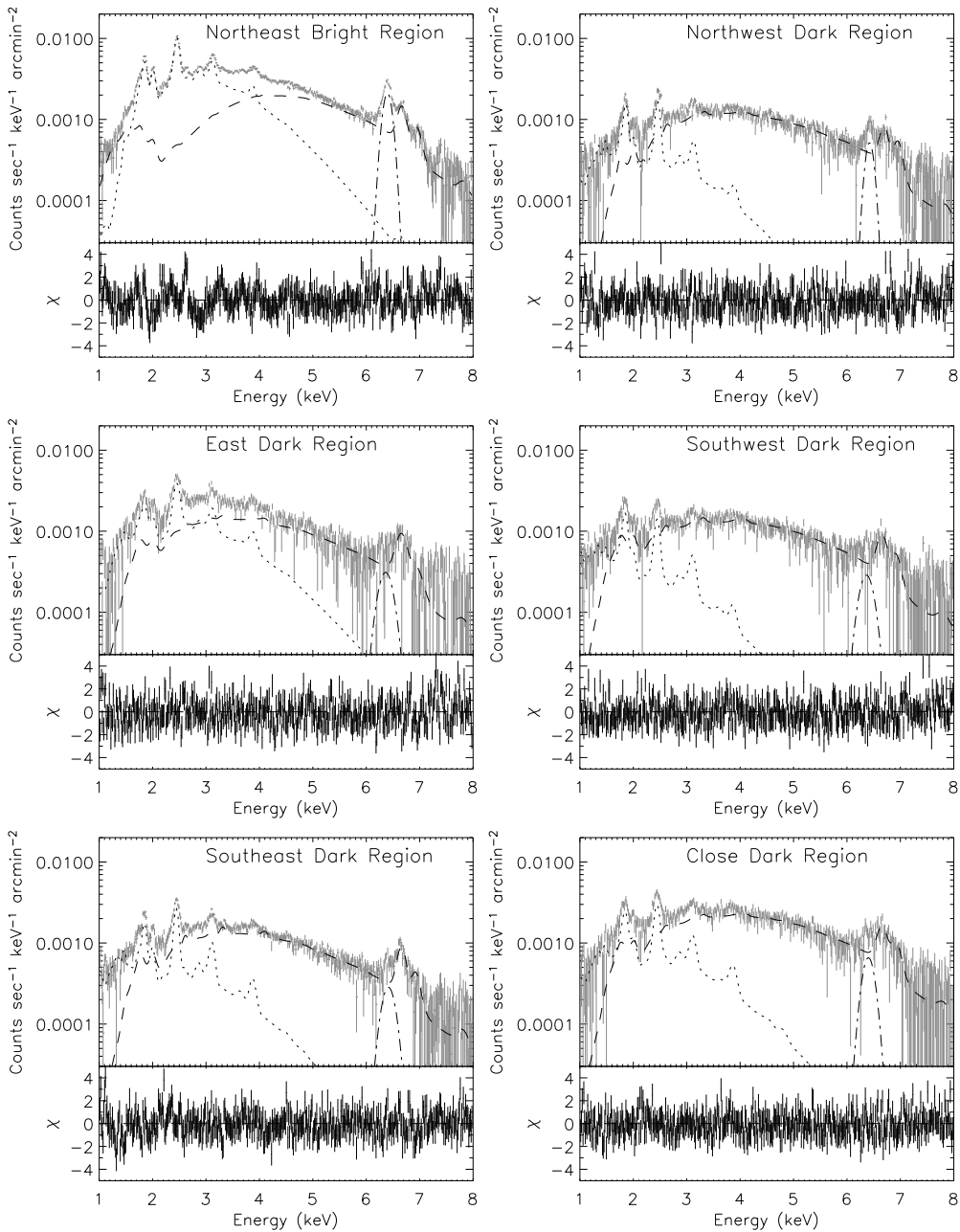,width=0.8\linewidth}}
\caption{
Background-subtracted spectra and best-fit two-$kT$ plasma model 
for each of the regions in Figure~2. The {\it dotted line} denotes
the soft $kT \approx 0.8$ keV plasma 
component, the {\it dashed line} indicates the hard $kT \approx 8$ keV 
plasma component, and the {\it dot-dashed line} indicates the 
low-ionization Fe line emission.}
\label{fig:diffmod}
\end{figure*}

Despite the complexity of the diffuse emission in the image, the spectra 
from all of the regions are basically described with the same 
collisional-equilibrium plasma model.
We list the best-fit parameters for this model in each of the regions in
Table~\ref{tab:twokt}, and display them in Figure~\ref{fig:diffmod}. The 
uncertainties are $1\sigma$, derived from a 
search in chi-squared space with $\Delta \chi^2 = 1.0$.
The value of $\chi^2$ from some regions is formally
poor, but this is not surprising given that the properties of the diffuse
emission probably vary somewhat within each of the regions in 
Figures~\ref{fig:softimg} and \ref{fig:hardimg}. 
We note that 
two-temperature non-equilibrium plasma models (\program{nei} and \program{vnei}
in \program{XSPEC}; Masai 1984) did not reproduce the data nearly as well, 
as they yielded $\chi^2_\nu > 1.5$ for all combinations of 
free parameters and absorption components. Moreover, the best-fit ionization
parameters under those models were larger than $10^{12}$ s cm$^{-3}$, which 
reinforces the suggestion that the diffuse emission originates from plasma 
in thermal equilibrium.

The absorption columns derived are consistent with the expected 
Galactic values. The soft, $kT \approx 0.8$~keV component has a total column 
of approximately 
$5\times 10^{22}$~cm$^{-2}$, assuming that the partial-covering absorber 
affects 95\% of the field. This indicates that only a small 
fraction of the soft diffuse emission originates in the foreground.
The hot, $kT \approx 8$~keV component is absorbed by about 
$4 \times 10^{22}$ cm$^{-2}$ over
the entire region, with additional absorption of $\sim 5 \times 10^{23}$
cm$^{-2}$ affecting on order half of each region. The additional 
partial-covering absorption column is comparable to that which 
could be provided by the typical molecular clouds in the region, 
$\sim 3\times 10^{23}$~cm$^{-2}$ (Zylka, Mezger, \& Wink 1990)\nocite{zmw90}. 
The higher absorption toward
the hard component is probably a selection effect, caused by the fact that 
no trace of the soft component could emerge from behind 
$5\times10^{23}$ cm$^{-2}$
of absorption. The differences in the absorption column from field-to-field  
are not surprising given the non-uniform distribution of 
molecular clouds near the Galactic center. 

The temperatures of the two plasma components,
$kT_{\rm cool} = 0.7-0.9$~keV and $kT_{\rm hot} = 6-9$~keV, both differ 
from region to region. 
The temperature of the cooler component
is well-constrained by the lines of Si, S, and Ar, and changes by at most
0.1~keV if a different absorption model is assumed. For example, assuming 
identical absorption for the cool and hot components, the derived temperature
decreases to 0.6~keV for some of the regions, but the differences persist.
Therefore, the dispersion in 
the values of
$kT_{\rm cool}$ are significant. On the other hand, the temperature of the hot 
component is determined both by the ratio of the H-like and He-like lines and 
by the continuum above 5~keV, and is more uncertain because the effective 
area of the detector there is smaller. Moreover, $kT_{\rm hot}$ differs by up
to 2~keV when the absorption model is changed. For instance, when a 
single absorption factor of the form of Equation~\ref{eq:abs} is assumed, 
$kT_{\rm hot}$ increases to 10~keV because the 
continuum is inferred to be flatter. As a result of the poorer constraint, 
the differences in $kT_{\rm hot}$ are only significant at the 
$2\sigma$ level.

The fluxes of both spectral components also vary spatially.
The emission measures are better correlated with the observed flux than 
are the temperatures, which suggests that the density or volume of the
plasma is the most
important factor determining the surface brightness of the diffuse emission.
The flux from both plasma components is highest in the East and Northeast, 
as expected from the 
images in Figures~\ref{fig:softimg} and \ref{fig:hardimg}. The differences in
the soft emission is the most dramatic: it is lowest in the Northwest, 
Southeast, and Southwest, 50\% higher in the 
region close to the Galactic center, a factor of 3 higher in the East, and 
a factor of 9 brighter in the Northeast. 
On the other hand, the observed flux from the hard 
component differs by less than 10\% in the four dark regions located at 
$\sim 8$\arcmin\ from \sgrastar, and is only a 60\% higher in the Close
region and in the Northeast. 
The inferred de-absorbed hard fluxes differ even less, as the brightest
regions produce only 25\% more flux than the darkest.

Under the two-temperature plasma model, there appear to be significant 
ranges in the elemental abundances both when comparing individual 
elements, and when comparing the dark and bright regions.
Unfortunately, we have found that the abundances of all of the metals can 
be made consistent
with solar values if we add a third thermal plasma component with 
$kT \approx 4$~keV to the model. Therefore, we must
view the measured abundances with skepticism. It will be necessary to 
obtain spectra of the 
diffuse emission with higher energy resolution to draw firm conclusions 
about the metal abundances in the emitting material. 

Finally, we note that the neutral Fe emission at 6.4~keV is a factor
of 4 more intense in the Northeast than in the rest of the regions examined.
This can be seen clearly in the equivalent width maps of \citet{par03}. The
strength of the 6.4~keV Fe emission clearly increases with that of the soft 
component of the diffuse emission in Figure~\ref{fig:softimg}, although it
tends to appear in more strongly localized regions than the soft plasma. 
In fact, neutral Fe emission is responsible for the bright filamentary 
features in the Northeast of the hard image \citep{par03}.

These models are the simplest that we have found that reproduce the data. 
However, the intrinsic spectrum conceivably could be more complicated. 
For instance, if the low-ionization Fe line is produced as part 
of a reflection nebula, it should be accompanied by a scattered continuum 
component \citep[e.g.,][]{mur00}. It has also been proposed that there
is a non-thermal component to the spectrum between 2--8 keV, which represents
a low-energy extension of the power-law emission seen above 10~keV 
\citep[e.g.,][but see Lebrun \etal\ 2004]{val00}. Adding a power-law
component with photon index $\Gamma = 1.8-2.5$ does not 
significantly improve the fit, although the lower bound on the temperature
of the hot plasma decreases to $\approx 6$ keV. Likewise, adding a third
plasma component does not improve the fit. None of the additional components 
change the basic conclusions that we present in Section~\ref{sec:disc}.

\subsubsection{Iron Emission\label{sec:iron}}

As we will discuss in Section~\ref{sec:disc}, a thermal plasma containing 
He-like and H-like ions of Fe will inevitably be 
too hot to be bound to the Galactic plane, and will therefore require
a very large amount of energy to sustain. This has led several authors to 
propose that much of the continuum and iron 
emission are produced by non-thermal processes. 
If this is the case, then the centroid energies, widths, and line ratios 
of the iron emission provide the best constraints on these models. Therefore,
we have modeled the Fe K-$\alpha$, He-$\alpha$, and 
H-$\alpha$ line complexes with three Gaussians. The Fe He-$\alpha$ line is 
a combination of forbidden (6.63 keV), inter-combination (6.67 keV), and 
resonance (6.70 keV) transitions that can not be resolved as separate lines
with the ACIS-I. Likewise, the Fe H-$\alpha$ line (6.97 keV) should contain
a contribution from the K-$\beta$ transition (7.03 keV), which has an 
intensity that is $\la 15$\% of that of the K-$\alpha$ transition
\citep[e.g.,][]{par03}. We modeled the continuum emission between
4.5--8.0 keV with a power law to facilitate the computation of equivalent 
widths. The resulting model parameters are listed in Table~\ref{tab:iron}.
The uncertainties are 1$\sigma$, derived from a search in chi-squared 
space with $\Delta \chi^2 = 1.0$ for the centroids, widths, and intensities,
and $\Delta \chi^2 = 2.3$ for the ratios of the intensities.

One of the best diagnostics of the non-thermal models are the centroid 
energies of the lines \citep[e.g.,][]{val00,mas02}.
Due to the shift in the gain of the ACIS-I mentioned above, 
the observed energies of the Fe He-$\alpha$ and H-$\alpha$ lines are 
shifted to lower energies by $\approx 0.5$\%. These shifts are 
significant at the $\la 2\sigma$ level. 
However, the Fe K-$\alpha$ lines lie, within their uncertainties, right at the 
expected energy of 6.40~keV. If we account for a 0.5\% shift to lower energies,
its true centroid could be as high as 6.43~keV. 

In some cases, non-thermal models for the iron emission also predict that the 
lines should be broadened \citep[e.g.,][]{tmh99}. The widths of the 
K-$\alpha$ and He-$\alpha$ lines can only be constrained
meaningfully with the two spectra that have the highest signal-to-noise, 
those from the Northeast and Southeast. In the Northeast, the Fe K-$\alpha$ 
line has a width of $39\pm7$ eV, while in the Southeast its width is $<70$ eV.
If the width in the Northeast is real, it would imply a velocity dispersion of 
$\approx 2000$~km s$^{-1}$. The Fe He-$\alpha$ line also appears to be 
resolved in the Northeast, with a width of $50\pm20$ eV, while in the 
Southeast the upper limit to the width is 40 eV. In both cases, however, the
width is consistent with the 70 eV separation between the recombination and 
forbidden lines. These widths are lower than those reported from ASCA
data by \citet{koy96} ($75 \pm 10$ eV). The constraints from the remaining
spectra are poorer, so we fixed the widths to 40~eV.

In all cases except the Northeast, the strongest of the iron lines is the 
He-$\alpha$ transition. It is a factor of 1.4--3.3 stronger than the 
K-$\alpha$ line, and a factor of 2.1--4.4 stronger than the H-$\alpha$ line.
However, the uncertainties on the line ratios are 
large enough that we can not identify significant variations between the
dark regions.
The Northeast region stands out with a very strong 
K-$\alpha$ line with equivalent width of 570 eV. For comparison, the 
Fe K-$\alpha$ features studied by \citet{par03} had equivalent widths of 
1~keV.

Overall, the properties of the iron emission are remarkably constant in 
the dark regions, with the only significant variations found in the bright
regions to the Northeast, and to a lesser degree in the East.

\subsection{Comparison to Point Sources\label{sec:ps}}

In this section, we examine three aspects of the point sources. First, 
we determine the amount of flux from detected point sources in each 
region, to evaluate whether the spatially varying detection
threshold affects the amount of flux attributed to diffuse emission.
Next, we compare the average spectrum of the detected point sources to that of
the diffuse emission. Finally, we assume that
\begin{figure*}[th]
\centerline{\epsfig{file=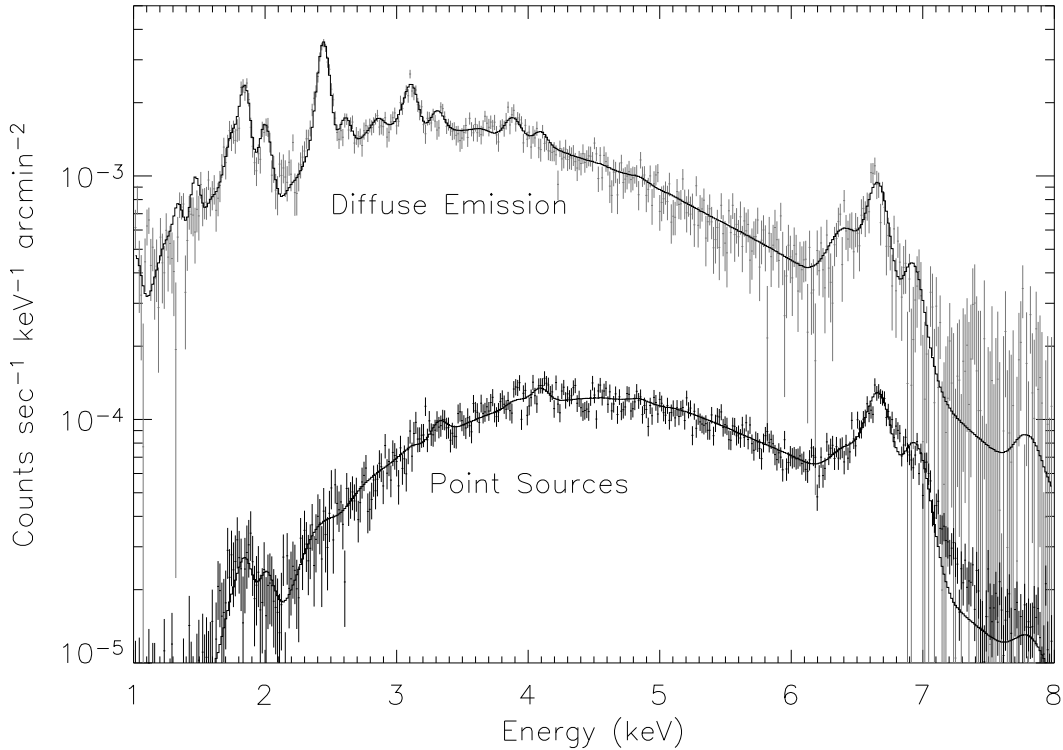,width=0.8\linewidth}}
\caption{
Background-subtracted spectrum of the diffuse emission from the southeast,
compared to the average spectrum of the detected point sources in the entire 
17\arcmin\ by 17\arcmin\ field. The spectrum of the diffuse emission has 
been normalized to the area from which it was extracted, while the spectrum
of the point sources has been normalized to an area of $17^2$ arcmin$^2$.
The detected point sources contribute no more than 20\% of the total
flux from the diffuse emission, even in the darkest regions of the image.}
\label{fig:psvdiff}
\end{figure*}
 undetected point sources have
spectra identical to the detected sources, and determine the maximum flux 
that they could contribute to the diffuse emission.

In Muno et al. (2004), we examine the spectra of the point 
sources 
in considerable detail; the methods for extracting and combining the spectra 
of the point sources are described there. In brief, we extracted spectra
from within the 90\% contour of the PSF around each point source
using the \program{acis\_extract} routine from the Tools for 
X-ray Analysis (TARA).\footnote{\html{www.astro.psu.edu/xray/docs/TARA/}}
We then summed the resulting spectra for each region. We computed the 
effective area
functions for each source using \program{mkarf}, corrected them for the 
fraction of the PSF enclosed by each region and 
for the hydrocarbon build-up on the
detectors. We then averaged the effective area weighted
by the count rate from each source. Finally, we averaged the response functions
that accompany the \citet{tow02b} CTI-corrector from the location of each 
point source, again weighted by the counts from each source. 
In order to confirm that the point-source spectra were not contaminated by 
diffuse emission, we re-extracted the spectra for several hundred point 
sources from regions that enclosed only 50\% of the PSF, and found that
the resulting average spectrum was indistinguishable from that extracted
from the 90\% contour of the PSF. 

Vignetting limits our ability to resolve point sources
at large offset angles from the aim-point, so we first established
that a failure to resolve point sources does not affect our estimates of the 
diffuse flux. To do so, we extracted a summed spectrum 
from the point sources in each region. We used the diffuse emission from 
each region to estimate the background, and modeled the point source
spectra with the same two-$kT$
model as for the diffuse emission. We list the surface brightness from 
detected point sources per square arcminute in Table~\ref{tab:difflux}, along
with the total, soft, and hard flux from the diffuse emission. The flux from
detected point sources is nearly identical in all of the dark regions located
$\approx 7.5$\arcmin\ from the aim-point. Therefore, we have probably been 
equally successful at resolving point sources in each dark region. A larger
flux from point sources is observed in the Northeast. Although this 
excess flux from point sources could indicate
that there are more stellar X-ray sources in this region, it is 
more likely that we mistakenly identified small knots in the diffuse 
emission as point sources. Larger knots that could be identified as extended 
by eye were removed from the point source list. The flux from detected point 
sources close to the Galactic center is higher because (1) the angular 
resolution is better within 5\arcmin\ of the aim point (\sgrastar), so 
that it is possible to detect fainter sources, and (2) the surface density 
of point sources increases as $\theta^{-1}$ toward \sgrastar.

Next, we compared the spectrum of the point sources from the entire field 
to that of the diffuse emission. 
We found in Muno et al. (2004) that the averaged 
spectrum of 
the point sources changed only slightly when considering sources with 
intensities
that ranged over a factor of 
\begin{figure*}[ht]
\centerline{\epsfig{file=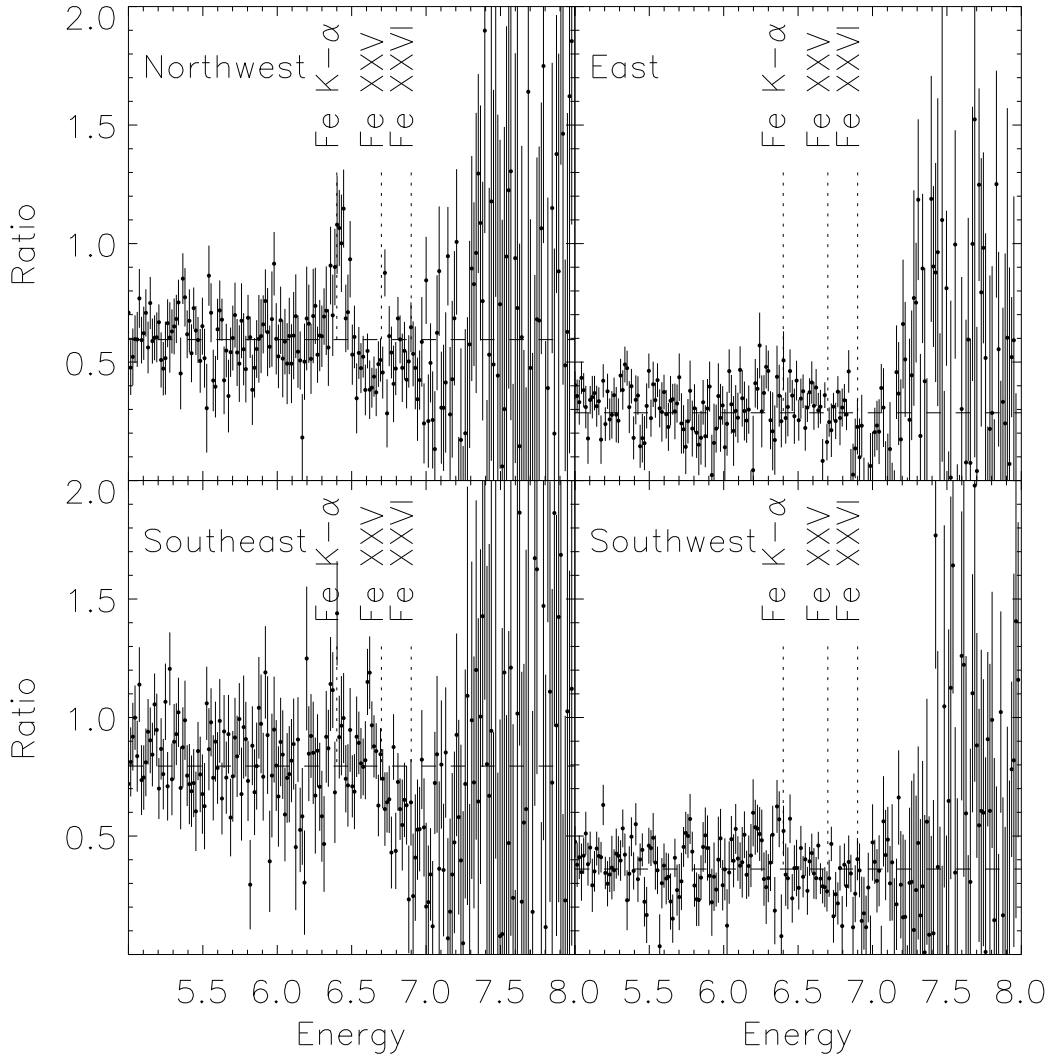,width=0.8\linewidth}}
\caption{
Ratios of the background-subtracted count spectra from diffuse emission
to that from faint point sources. Ratios are displayed for the four dark
regions located 7\arcmin\ from \sgrastar. The dashed line indicates the 
mean ratio between 5 and 7 keV. The ratio is nearly
constant, which indicates that the spectral shapes are very similar. 
We note that the close region yields similar 
results, while the ratio from the Northeast exhibits significant deviations
due to strong low-ionization iron emission.}
\label{fig:pscomp}
\end{figure*}
$\sim 20$,  so we chose to compare the
spectrum of point sources with fewer than 80 net counts to that of 
the diffuse emission. 
We used the spectrum of the diffuse emission 
from the entire field to estimate the background for the point sources.
The average spectrum of the point sources and spectrum of the diffuse emission
from the southeast are displayed 
in Figure~\ref{fig:psvdiff}, while 
the ratios between count spectra
from the dark regions of diffuse emission to that from the point sources
is displayed in Figure~\ref{fig:pscomp}.
As noted by \citet{wgl02}, the shapes and relative intensities of the 
He-like and H-like iron lines appear qualitatively similar in the 
point sources and diffuse emission. 
Under the two-$kT$ plasma model for the point sources, we find 
that the soft 0.8~keV component is heavily absorbed, so that it contributes 
only 3\% of the total observed 2--8 keV flux. The hot 8~keV plasma component
produces the remainder of the observed flux. Therefore, the near-absence
of a soft component is the main reason that the point source spectra appear
much harder.

In order to determine an upper limit to the amount of flux that undetected
point sources can contribute to the diffuse emission, we have added
a spectral component representing the point sources with fewer than 
80 net counts to our two-$kT$ models for the diffuse emission. 
Only a constant normalization for the fiducial point source spectrum was
allowed to vary, while the parameters of the spectral components representing
the diffuse emission were allowed to vary as in Section~\ref{sec:twokt}.
We increased the value of the constant 
normalization until $\chi^2$ exceeded a threshold that corresponded to a 
10\% chance that the model and data were consistent. If the 
initial model was acceptable, the threshold was $\chi^2 = 506$ for 456 
degrees of freedom. However, in a couple of cases the initial model including
the point source spectrum was unacceptable at the 90\% confidence level (see
Table~\ref{tab:twokt}),
so we varied the normalization until $\chi^2$ increased by 
$\delta\chi^2 = (506-456) \cdot \chi^2/456$ (this is equivalent to inflating 
the assumed uncertainties to force $\chi^2_\nu$ to equal 1 for the best-fit 
model). 

The surface brightness of the diffuse emission that can be accounted for 
by undetected point sources is listed by region in Table~\ref{tab:difflux}.
Undetected point sources may produce $(0.7-1.5) \times 10^{-13}$
\ergcmsarcmin\ in the regions 7\arcmin\ from the Galactic center, which 
is 35--80\% of the total observed diffuse flux. The upper limits on 
the fluxes from undetected point sources are 2--5 times larger than the 
total flux from the observed point sources. We also list in 
Table~\ref{tab:difflux} the flux from the
soft and hard components that remains when an undetected point source 
contribution is subtracted. 
The reduction 
in inferred diffuse flux is most dramatic in the hard band: in the 
Southeast, Southwest, and Close dark regions, a hard diffuse component 
is not necessary if one includes a contribution of undetected point sources 
in the spectrum. Moreover, the remaining three regions in which undetected 
point sources cannot replace the hard component of the diffuse emission 
are also those regions in which the initial model for the diffuse emission
(without the point source component) was formally unacceptable. 

The similarity between the spectra of the point sources and diffuse 
emission suggests that all of the hard diffuse emission could be produced
by point sources. However, in this case the larger hard flux in the 
Northeast would require that there are $\approx 40$\% more stellar X-ray 
sources per solid angle than in the dark regions at a 
similar offset from the Galactic center. We address this issue further in 
Section~\ref{sec:point}, 
where we estimate the number of undetected sources that would be required 
to produce the observed $kT\approx8$ keV diffuse emission.

\section{Discussion\label{sec:disc}}

As mentioned in the introduction, these \chandra\ observations of the 
diffuse emission within the 17\arcmin\
by 17\arcmin\ field around \sgrastar\ provide three advantages over 
previous observations with \asca\ \citep{koy96} and \bepposax\ \citep{sm99}: 
(1) the instrument does not produce line
emission between 2--7 keV that could contaminate the spectrum and thereby
produce ambiguities in the measured ionization state of the emitting
medium \citep[compare Figure~\ref{fig:diffraw} and, e.g.,][]{tan02}, 
(2) the long integration time allows us to measure the spectrum of 
the diffuse emission with high signal-to-noise on arcminute spatial scales
(Figure~\ref{fig:diffmod}), 
and (3) the high angular resolution of \chandra\ 
allows us to distinguish 
the truly diffuse emission from point sources, supernova remnants, and 
an apparent outflow from \sgrastar\ (Figure~\ref{fig:rawimg}). 

The spectrum of the diffuse emission from the Galactic center is dominated by 
line emission from He-like and H-like Si, S, Ar, Ca, and Fe 
\citep[Figure~\ref{fig:diffraw}; see also][]{yam97, tan02}. These lines 
indicate that the diffuse emission results from a two-component 
collisionally-ionized plasma, with temperatures of 
$kT_{\rm cool} \approx 0.8$~keV and $kT_{\rm hot} \approx 8$~keV. 
The line energies and ratios are consistent
with those expected from plasma in thermal equilibrium
\citep[compare][]{kan97,tan02}. Therefore, in Section~\ref{sec:prop}, 
we adopt the assumption of thermal equilibrium to examine the physical 
properties of the putative plasma responsible for the diffuse emission.

The main parameters of our spectral models, the temperature and emission 
measure of the putative plasma components, vary on scales of $\approx 10$~pc. 
The properties of the cool, $kT_{\rm cool} \approx 0.8$ keV component were 
determined 
primarily by the ratio of fluxes in the He-like and H-like transitions of 
Si and S (Figure~\ref{fig:diffmod}), and were well-constrained. The
temperature of the soft component of the plasma ranges 
between 0.7 to 0.9~keV, and
its emission measure ranges between 1--24 cm$^{-6}$ pc (Table~\ref{tab:twokt}).
The properties of the hot, $kT_{\rm hot} \approx 8$ keV component 
were derived from the ratio of fluxes in the He-like and H-like transitions 
of Fe, and by the shape of the continuum. 
This hard component is less well-constrained, because the instrumental
effective area is much smaller near the Fe transitions at 6--7 keV than it is
near the Si and S transitions at 1.5--2.5 keV. The temperature
of the hard component ranges between 6--9~keV, but is consistent with a mean 
value of 8~keV at the 2$\sigma$ level. The emission
measure ranges between 1.5--3.0 cm$^{-6}$ pc (Table~\ref{tab:twokt}). 
These spatial variations in the diffuse emission provide 
new insight into the origin of the putative plasma components that produce
the diffuse emission, as we discuss in 
Sections~\ref{sec:disc:soft} and \ref{sec:disc:hard}.

Finally, we were able to  
separate the diffuse emission from point sources as faint as 
$3\times10^{-15}$~\ergcms. We have already noted that the flux from the 
point sources detected in our \chandra\ image accounts for less than 
10\% of the diffuse emission from the Galactic center \citep{mun03}. 
However, in the current paper we find that the average spectrum of the 
faintest point sources detected near \sgrastar\ is remarkably 
similar to that of the hard component of the diffuse emission 
\citep[see also][]{wgl02}.
The similarity is particularly striking between 6.5--7.0~keV, where
there is strong emission from He-like and H-like Fe 
(Figure~\ref{fig:psvdiff}).
As a result, if point sources that have not been detected have the same spectra
as the detected ones, they could contribute up to 80\% of 
the total 2--8~keV flux, and up to 100\% of the hard component of the 
flux (Table~\ref{tab:difflux}). We address
the plausibility that undetected point sources produce a significant 
fraction of the apparently diffuse flux in Section~\ref{sec:point}.

\subsection{Plasma Properties\label{sec:prop}}

The two-$kT$ plasma model provides as free parameters the temperature ($kT$)
and emission measure ($K_{\rm EM}$) of the plasma components.
By assuming a depth for the emitting region, we can use these parameters
to derive energies, densities, and time scales that can be used to 
understand the origin of the putative plasma. 
We list the properties for the putative soft and hard plasma components 
in Table~\ref{tab:prop}. We assume a distance of 8 kpc to the 
Galactic center throughout the rest of the paper \citep{mcn00}.

The luminosity of the diffuse emission provides a lower limit to the 
amount of energy required to sustain it. We have computed the 
luminosity from the de-absorbed 2--8 keV fluxes in Table~\ref{tab:twokt}
by applying a bolometric correction that we determined by integrating 
the total flux from the \program{vapec} model in 
\program{XSPEC}; this is equivalent to using the cooling function in, 
for example, \citet{ray76}. The bolometric correction for the soft component 
is $\approx 20$, and for the hard component is $\approx 2$. Most of the extra 
luminosity lies between 0.1--2~keV, and would be easily detectable with 
\chandra\ were it not for the absorption toward the Galactic center. 
These corrections are uncertain by about 50\%, as
they depend upon the assumed elemental abundances.  
 With these caveats in mind, we find that 
the luminosities of the soft component range between 
$(6-12)\times 10^{33}$~\ergsarcmin\ in the darkest regions, up to 
$9\times 10^{34}$~\ergsarcmin\ in the Northeast bright region. The 
luminosity of the hard component spans a smaller range, from 
$(5-9)\times 10^{33}$~\ergsarcmin. 

The variations in the luminosity of the plasma are correlated with those 
in the emission measure, $K_{\rm EM} = \int n_e n_H dV$, and 
therefore either the depths 
of the emitting regions or the densities of the plasmas vary 
spatially over the image in Figure~\ref{fig:twocolor}. 
The depths of the emitting regions are probably somewhere between 
10 pc, which corresponds to the approximate diameter of the extraction 
regions we used, 
and 250 pc, which corresponds to the $1.8^\circ$ major axis of the elliptical 
region of 6.7~keV Fe emission region observed with {\it Ginga} \citep{yam90}. 
We will take the geometric mean of these extreme values, and assume that the 
depth is 50 pc. We report each quantity in Table~\ref{tab:prop} with a 
scale factor $d_{50}$ to account for possible variations in the 
depth of up to a factor of 5. This term can also account for the possibility 
that the filling 
factor of the plasma is less than unity. Because we are only 
interested in order-of-magnitude estimates, we assume that the plasma is 
pure hydrogen. The density of the plasma is then related to the emission 
measure by $n_e = 0.1 d_{50}^{-1/2} K_{\rm EM}^{1/2}~{\rm cm}^{-3}$.
The mean density of the 
soft plasma is 0.1 cm$^{-3}$ in the darkest regions, and 0.5 cm$^{-3}$ in 
the bright region to the Northeast. The density of the hard plasma is near
0.1 cm$^{-3}$ over most of the image, and increases by a factor of 2 in the
Northeast and within 4\arcmin\ of \sgrastar. For comparison, 
\citet{koy96} derive densities of 0.3--0.4 cm$^{-3}$ from their analysis
of a larger, 1 square degree field observed with \asca. This is probably
because the larger field is dominated by bright regions similar to the 
Northeast region in our image, such as the Radio Arches region in the
survey of \citet{wgl02}.

From the density and volume of the plasma, we can compute its total mass.
The masses of both the soft and hard components in the dark regions are 
about $1 d_{50}^{1/2}$~\msun\ arcmin$^{-2}$. The density of 
soft plasma in the bright 
region is higher, so its mass is $3 d_{50}^{1/2}$~\msun\ arcmin$^{-2}$. The
total mass of the plasma in the 17\arcmin\ by 17\arcmin\ field is about
500 $d_{50}^{1/2}$ \msun. For comparison, in the 1 square degree \asca\ 
field, \citet{koy96} derive a plasma mass of 2000--4000~\msun, which is 
only slightly lower than our mass if one takes into account the difference in
field-of-view.

The energy density of the plasma is $U = \frac{3}{2} nkT$, 
and so is proportional to  $d_{50}^{-1/2}$.
In the dark regions, the soft component has an energy density of 
$3\times10^{-10}$ erg cm$^{-3}$, or 200 eV cm$^{-3}$. The hard component
has an energy density of $1\times10^{-9}$ erg cm$^{-3}$, or 1 keV cm$^{-3}$.
The total energy per arcmin$^2$ is found by multiplying
$U$ by the plasma volume, and so is proportional to $d_{50}^{1/2}$.
In the dark regions, the soft component has 
$E = 2\times10^{48}$~erg arcmin$^{-2}$, while the hard component has 
$E = 2\times10^{49}$~erg arcmin$^{-2}$. These values are nearly identical
to those derived from the \asca\ observations of \citet{koy96}. In the 
bright regions, the soft 
component has three times as much energy as it does in the dark regions, 
while the hard component contains only
50\% more energy. The total thermal energy of the plasma in the image is over
$10^{51.8}$~erg.

Dividing the total energy of the plasma by its luminosity yields the
cooling time scale, $t_{\rm rad}$. In the dark regions, the soft 
component cools in $6\times10^6 d_{50}^{3/2}$ y. The bright region 
should cool more rapidly, in $3\times10^6 d_{50}^{3/2}$ y, because 
$t_{\rm rad} \propto n^{-1}$. The hard component cools in 
$1\times10^7 d_{50}^{3/2}$ y.

However, it has been previously noted that an 8~keV plasma is too hot to 
be bound to the Galactic plane \citep[e.g.,][]{yam90,kan97, yam97}, 
so it is important
to consider the amount of energy that could be lost as the plasma expands. 
If we use the Galactic potential from 
Breitschwerdt, McKenzie, \& V\"{o}lk (1991)\nocite{bmv91},
we find that the escape velocity from the Galactic center ($R \approx 20$ pc)
is approximately 900 km s$^{-1}$. This can be compared to the sound speed of 
the plasma 
$c_s = (\gamma kT/\mu m_p)^{1/2}$. If we assume $\gamma = 5/3$ 
(for a monotonic, adiabatic gas) and $\mu=0.5$ (electrons and protons), the 
sound speed for the 0.8 keV plasma is 
$\approx 500$ km s$^{-1}$, and for the 8 keV plasma is $\approx 1500$~km 
s$^{-1}$. Therefore, only the cooler plasma is gravitationally bound to 
the Galaxy. However, in the absence of other confining forces, even
the soft plasma could 
expand significantly; the contribution of the 
$\approx 3\times10^7$~\msun\ within the central 10 pc of the Galaxy
(Laundhardt, Zylka, \& Mezger 2002)\nocite{lzm02} is 
negligible when considering whether the plasma can expand, as the 
corresponding escape velocity from the central parsecs is only 150 km s$^{-1}$.
Using the potential in \citet{bmv91}, the 0.8 keV plasma could
expand to a height of several hundred parsecs.

Upper limits to the energy required to sustain the plasma components can 
be obtained by assuming that they expand adiabatically and form a wind
from the Galactic center. Although computing the energy loss rate rigorously
would require a full solution of wind equations such as those in 
\citet{cc85}, we can make a rough estimate of it by assuming 
$\dot{W}_{\rm exp} = P dV/dt \approx P V^{2/3} c_s$. 
For the soft component of the
plasma, the energy loss rate is $\sim 3\times10^{36}$~\ergsec arcmin$^{-2}$, 
300 times higher than the X-ray luminosity of the plasma.
For the hot component of the plasma,
this energy loss rate is $\sim 1\times10^{38}$~\ergsec arcmin$^{-2}$, or 4 
orders of magnitude higher than its X-ray luminosity. Over the entire image, 
the upper limit to the power required is
$\sim 10^{40}$ erg s$^{-1}$. The corresponding cooling time if the 
plasma is expanding is $\sim 2\times10^{4}$ years for the $kT \approx 0.8$~keV
plasma, and $\sim 6\times10^3$ years for the $kT \approx 8$ keV plasma.
These are much shorter than the radiative cooling time scales.

Even if external forces are on average sufficient to confine the diffuse 
plasma, any over-density in the plasma should be smoothed out in a short 
time by the differential rotation at the Galactic center. 
At 10 pc and for an enclosed mass of $3\times10^7$~\msun\, 
the orbital time scale is $t_{\rm orb} \approx 6\times10^5$ years.
Therefore, differential rotation would smooth out any variations in the 
plasma properties with a radial extent $\Delta R$ on a time scale of 
$t_{\rm tidal} \approx  (R/\Delta R) t_{\rm orb}$. For parsec-scale features,
$t_{\rm tidal} \approx 6 \times 10^{6}$ years, which is comparable to 
the cooling time of the soft plasma, but significantly shorter than the 
cooling time of the hard plasma.

\subsection{Soft Component\label{sec:disc:soft}}

The pronounced variations in the surface brightness of the soft, 
$kT \approx 0.8$~keV component of the diffuse emission 
(Figure~\ref{fig:softimg}) have important 
implications for the spatial distribution and age of the putative 
plasma that produces it. The fact that the spatial variations are much 
more pronounced in the soft emission than in the hard (Table~\ref{tab:prop})
suggests that the soft plasma occupies a much smaller volume. Indeed, if 
we assume that the soft and hard plasma are in pressure 
equilibrium, then the filling factor of the soft plasma would have to 
be roughly 1--10\% of that of the hard plasma. At the same time, any 
over-density in the soft plasma is unlikely to survive very long, because
the differential rotation of the Galactic center should shear apart any
coherent features. For example, the bright diffuse emission 
in the Northeast has an angular extent that corresponds to a size of 
$\approx 20$ pc, and so differential rotation should dissipate it within 
the orbital time scale of $6 \times 10^5$ y.  

The youth and small filling factor of the $kT \approx 0.8$ keV plasma
are easily understandable if it is produced by supernova remnants. This 
is the common explanation for the origin of similar soft diffuse emission
that is observed throughout the Galactic disk,
because supernova remnants are often 
observed to have spectra consistent with $\sim 1$ keV plasma 
\citep[e.g.,][]{kan97}.
Supernova are also the largest known source of energy for heating the ISM 
\citep{schl02}, and can easily provide enough energy to heat the diffuse
soft plasma. If we assume that the dominant cooling mechanism for the soft
plasma is radiative, then the energy input required to sustain it is 
only $\approx 3\times10^{36}$ erg s$^{-1}$ for the inner 20 pc of the Galaxy.
If, on analogy with the conclusions of \citet{kan97} for the Galactic disk, 
we assume that $\sim 1$\% of the $10^{51}$ erg of kinetic energy per 
supernova heats the soft plasma in our image, then it could be 
sustained by supernova occurring at a rate of one every $10^5$ y. 
This is not unreasonable, because the total Galactic supernova rate is thought 
to be on order 1 per 100 y, and the inner 20 pc of the Galaxy contains 
approximately 0.1\% of the Galactic mass \citep{lzm02}, so the expected
rate in this field is also about one supernova per $10^5$ y.
Moreover, Sgr A East \citep{mae02} and the radio wisp 'E' \citep{ho85} are 
already 
thought to be remnants of recent supernova, so it seems likely 
that the inner 20 pc of the Galaxy has experienced a supernova rate at
least this high. 

It is also possible that the winds from young, massive Wolf-Rayet and early O 
stars could contribute to the $kT \sim 0.8$ keV plasma. 
A typical WR star can lose mass at a rate of 
$\dot{M} \approx 10^{-5} M_\odot$ y$^{-1}$ with a velocity of 
$v \approx 2000$ km s$^{-1}$ \citep[e.g.][]{lrd92,sh03}. The kinetic 
energy of such a wind is $3 \times 10^{37}$~\ergsec.
X-rays are produced by internal shocks in the winds of individual stars,
but diffuse X-rays are only produced by the large shocks that 
occur when winds from clusters of these stars encounter the interstellar 
medium. 
In observations of massive star clusters, about 10\% of the wind 
kinetic energy is converted into X-rays \citep{sh03,tow03}. 
Therefore, a single WR and/or early O star
could in principle produce the soft component of the diffuse emission.
However, extended X-ray emission from massive stars is usually only 
associated with young stellar clusters  
that contain several massive stars within a core radius of $\sim 2$~pc,
which typically produce X-rays in a region only $\sim 3$ pc in radius 
\citep{tow03}.  
Therefore, the X-ray emission from the winds of massive stars is likely 
to only be important within an arcminute of any as-yet-undiscovered
star clusters.

The detection of strong $kT \approx 0.8$ keV emission to the Northeast of the
Galactic center is also consistent with our assumption that the soft plasma
originates from supernovae and winds from massive stars. The
Northeast region possesses several interesting 
properties that may be related to the enhanced diffuse X-ray emission: 
it contains large amounts of molecular gas 
\citep[][Tsuboi, Handa, \& Ukita 1999; Mezger, Duschl, \& Zylka 1996;]{dah98}\nocite{tsu99,mdz96}, it is located 
between young, massive star clusters at the Galactic center 
\citep{kra95,pau01} and the Radio arches region \citep{fig99}, 
and it is the site of the strongest low-ionization Fe emission
\citep{par03}. The mere presence of molecular clouds is clearly 
not sufficient for forming the soft diffuse and 6.4~keV Fe
emission, as  molecular clouds are also evident in the Southeast, without 
any corresponding enhancement in the iron emission \citep{par03}. 
Given the arguments for the origin of the soft plasma above, it is more
likely that both 
the soft diffuse emission and the neutral Fe emission are associated with 
recent star formation. For instance, a Type~II supernova in the Northeast 
could produce both the bright, soft diffuse emission and the neutral Fe 
emission \citep[e.g.,][]{mae02,byk02}. Outside of the
Galactic center, the most similar region may well be the Carina Nebula,
which exhibits $\approx 2\times10^{35}$~\ergsec\
of X-rays from the outflow around 
$\eta$ Car, the WR and O stars in the cluster Trumpler 14, and diffuse 
emission that may have resulted from recent supernovae \citep{sc82}. 

\subsection{Hard Component\label{sec:disc:hard}}

The hard, $kT \approx 8$~keV emission is distributed much more uniformly 
than the soft, but its intensity is still correlated with that of the 
soft emission. The correlation between the hard and soft emission
suggests that they are produced by related physical processes. The 
relative uniformity of the hard emission may result from its higher 
sound speed, which would cause over-dense regions of $kT \approx 8$~keV
plasma to expand on a time scale of $\sim 10^4$ y.

However, the $kT \approx 8$ keV plasma is somewhat hotter than is usually 
observed from either supernova remnants or clusters of WR and early O stars
\citep[see, e.g.,][]{mae02,tow03}.
Moreover, the sound speed of the hot plasma
is larger than the escape velocity from the Galactic center, and therefore
the energy required to sustain the expanding $kT \approx 8$ keV plasma 
within the image in Figure~\ref{fig:twocolor} is $\sim 10^{40}$ erg s$^{-1}$.
This energy is four orders of magnitude larger than that required to 
sustain the $kT \approx 0.8$~keV plasma (which probably does not cool 
by expanding), and is equivalent to the entire kinetic energy of one supernova 
occurring every 3000 y. This makes it difficult to understand the origin of 
this putative hot plasma, because supernovae are assumed to be the largest 
source of heat for the ISM. Moreover, using the values in 
Table~\ref{tab:prop}, the plasma flowing out from within the inner $R = 20$ pc
of the Galaxy would carry away a mass of roughly 
$\dot{M} \sim 4\pi R^2 m_p n c_s \sim 10^{-2}$ \msun\ y$^{-1}$. 
This mass loss rate is also equivalent to that from one supernova 
occurring every 3000 y. Finally, this $kT \approx 8$ keV diffuse emission 
is observed 
from throughout the Galactic disk, so the mechanism producing it must be 
widespread. In particular, the hot diffuse emission is probably not produced
by the super-massive black hole at the Galactic center, \sgrastar.

Several mechanisms have been proposed to explain the origin of the 
$kT \approx 8$ keV component of the Galactic diffuse emission. 
One possibility is that the hard plasma is heated by magnetic reconnection
that is driven by the turbulence that supernovae generate in the ISM 
\citep{tan99}. Magnetic connection could heat the plasma to 
$nkT \sim B^2/8\pi$, or, for $n\approx 0.1$ cm$^{-3}$
and $B \approx 0.2$ mG (Table~\ref{tab:prop}), $kT \sim 8$ keV. Fields 
of comparable strength are inferred to exist near the Galactic center 
\citep[e.g.,][]{sm96}. Moreover, \citet{tan99} also pointed out that 
with the right geometry, the same fields would also be strong enough to 
confine the $kT \approx 8$ keV plasma, thus possibly reducing the amount of 
energy required to sustain it. Unfortunately, the magnetic fields toward 
the Galactic center appear unlikely to confine the hot 
plasma. Individual magnetic flux tubes are observed as 
synchrotron-emitting radio filaments oriented perpendicular to the Galactic
plane \citep{yzm87,lar00}. These filaments seem to be interacting with
molecular clouds in the region, and are therefore thought to have pressures 
comparable to those of the turbulent molecular clouds, so that 
$B \approx 1$ mG \citep[compare][]{sm96,dah98}. However, it is not
clear whether the filaments represent only a small fraction of vertical fields 
that pervade the Galactic center (Serabyn \& Morris 1996; Chandran, Cowley, 
\& Morris 2000)\nocite{sm96,cha00}, or whether 
they represent purely local magnetic features \citep{che92,lar00b, yz03}.
There is also a toroidal component observed through polarization measurements
that dominates within 
molecular clouds \citep{nov03}. However, the Zeeman-splitting measurements of 
\citet{ug95} placed upper limits of $\approx 0.3$ mG to 
the strength of any arcminute-scale ordered fields along the lines-of-sights 
towards most of the molecular clouds near the Galactic center; the
only Zeeman measurements that reveal fields $\ga 1$ mG are toward the 
circum-nuclear disk \citep{klc92}. Taken together,
these observations suggest that the magnetic fields are
predominantly vertical at the Galactic center, with a toroidal component 
produced by orbital shear in the molecular clouds \citep{ug95,cha00,nov03}. 
Therefore, any hot plasma at the 
Galactic center would be able to escape vertically away from the plane, 
thus forming a wind or fountain of plasma \citep[e.g.,][]{bmv91}. 

Another class of hypotheses assume that the hard component of the Galactic 
diffuse emission is produced by non-thermal processes associated with 
supernova shocks. For instance, low-energy ($< 1$ MeV) 
cosmic rays could interact with the neutral ISM to produce
continuum emission through bremsstrahlung radiation and line emission through 
charge-exchange interactions \citep[e.g.,][]{vm98,tmh99,val00}. The cosmic-ray 
electrons could be accelerated by supernova or young pulsars 
\citep[e.g.,][]{schl02}. Originally, this model 
was attractive because it could also explain the non-thermal X-ray emission 
observed above 10~keV. However, observations with {\it INTEGRAL} 
have resolved $\approx 85$\% of the emission above 20~keV into discrete point 
sources \citep{leb04}. Moreover,
neither non-thermal electrons nor protons are efficient at producing 
bremsstrahlung radiation, so the energy required to produce the observed
diffuse emission is nearly as large as that required to replenish a 
continually-expanding thermal plasma \citep{dog02}. Finally, the model
presented by \citet{val00} does not predict the presence of the Fe H-$\alpha$
line at 6.9 keV, which is clearly evident in our \chandra\ spectra 
(Figure~\ref{fig:diffraw} and Table~\ref{tab:iron}).

Alternatively, the hard diffuse emission could originate from a quasi-thermal 
plasma that is generated by supernova shocks that propagate through the 
cooler 0.8~keV plasma \citep{dog02,mas02}. This process is a 
factor of $\sim 30$ more efficient at generating X-rays than is an expanding
thermal plasma.  Therefore, an energy input of $5 \times 10^{38}$~erg s$^{-1}$,
or the kinetic energy of one supernova occurring every 
$7\times10^4$ y, is required to produce the observed hard diffuse emission. 
However, neither of the two model spectra presented by \citet{mas02} are 
entirely consistent with the data in Table~\ref{tab:iron}. 
Their models predict that Fe 
K-$\alpha$ lines should be of comparable strength as the He-$\alpha$ line 
at 6.7~keV, whereas in all of the dark regions the Fe K-$\alpha$ 
lines are a factor of 2--3 weaker, and in the Northeast bright region it is 
a factor of 1.8 stronger. Their models also predict that
the Fe He-$\alpha$ lines should be a factor of 1.7--2.5 stronger than the 
Fe H-$\alpha$ lines (depending upon whether the background plasma has a 
temperature of 0.6 or 0.3 keV), whereas we find values that are 
generally higher, albeit only at the 1--2$\sigma$ level in each case. Finally, 
they predict that if the shocks form 
in a cool (0.3~keV) background plasma, then the low-ionization Fe line should 
have a centroid near 6.5~keV. The observed centroids are consistent with 
6.4~keV, and values as high as 6.5~keV are excluded at the $\ga 6\sigma$ level
in the Southwest and Northeast.
Therefore, the non-thermal
models that have been published to date are challenged by 
the remarkable similarity
between the observed spectrum and that expected from a $kT \approx 8$~keV
plasma in thermal equilibrium. However, further exploration of the 
parameter space of the non-thermal models is needed to confirm or refute them
definitively.


\subsection{The Number of Undetected Point Sources\label{sec:point}}

The spectrum of the $kT \approx 8$~keV plasma is also very similar to that 
of the faintest detected point sources. Therefore, in this section we use the 
$\log(N) - \log(S)$ 
distribution from Equation 5 in \citet{mun03} to compute the number of 
point sources that would be required to produce the hard diffuse emission in
the darkest region of the image: $1.7 \times 10^{-13}$ \ergcmsarcmin\ in the
Northwest. The $\log(N) - \log(S)$ distribution allows 
us to compute the surface density of sources down to a limiting flux 
$S$. Adapting the results of \citet{mun03}, we find:
\begin{equation}
 N(S) = \left\{ \begin{array}{ll}
 {{20} \over {\theta}} \left( {{S} \over {3\times 10^{-15}}} \right)^{-1.7\pm 0.2} &
S < 6 \times 10^{-15} \\
{{5} \over {\theta}} \left( {{S} \over {6\times 10^{-15}}} \right)^{-1.34\pm 0.08} &
S > 6 \times 10^{-15}.
\end{array} \right.
\label{eq:mod}
\end{equation}
We have made several modifications to this equation from the original version. 
First, we have included a 
factor $\theta$ for the offset from \sgrastar\ in arcminutes, which accounts 
for the decrease in surface 
density of the point sources within 9\arcmin\ from the Galactic center. Second,
we have converted the photon fluxes in \citet{mun03} to energy fluxes by 
assuming 1~\phcms~$= 8\times10^{-9}$~\ergcms\ (2.0--8~keV), which is 
appropriate for a $\Gamma = 0.5$ power-law spectrum absorbed by a
$6 \times 10^{22}$~cm$^{-2}$ column of gas and dust.
Finally, we have normalized 
the distribution in Equation~\ref{eq:mod} to the surface density at 
$\theta = 1$\arcmin,
whereas Equation~5 from \citet{mun03} was normalized
to the density at 4.5\arcmin. We can then obtain the total flux from
point sources  ($F_{\rm ps}$) in any given region by integrating the 
power laws in Equation~\ref{eq:mod}:
\begin{eqnarray} 
\nonumber F_{\rm ps}(> S_{\rm min}) = \\
\nonumber N_o S_o {{\alpha} \over {\alpha-1}} \left[
\left( {{S_{\rm min}} \over {S_o}} \right)^{-(\alpha - 1)} - 
\left( {{S_{\rm max}} \over {S_o}} \right)^{-(\alpha - 1)} \right]
\\
\times \int {{1} \over {\theta}} dA.
\label{eq:flux}
\end{eqnarray}
Here, $\alpha$ is the power-law slope, $S_o$ is the scale factor for the 
flux, and $N_o$ is the normalization of the power law, all from 
Equation~\ref{eq:mod}. $S_{\rm min}$ and $S_{\rm max}$ are the bounds on 
the point-source flux over which the total flux $F_{\rm ps}$ is computed. For 
the bright end of the distribution in Equation~\ref{eq:mod}, we 
assume $S_{\rm max}$ is $4\times 10^{-13}$~\ergcms, which is the brightest
point source that we observe in the image; the values of $F_{\rm ps}$ are 
not very sensitive to this assumed upper bound. The number of point sources
in a region is then found by inserting the assumed value of $F_{\rm ps}$
and solving for $S_{\rm min}$ in Equation~\ref{eq:flux}, and then inserting 
$S_{\rm min}$ into Equation~\ref{eq:mod}. Based on the observed hard diffuse
and point source flux in the Northwest (Table~\ref{tab:regions} and 
\ref{tab:difflux}), 
we take $F_{\rm ps} = 4\times10^{-12}$~\ergcms, and find 
$S_{\rm min} = 1 \times 10^{-16}$~\ergcms. Using this limiting flux 
in the $\log(N)-\log(S)$ distribution, we would predict a surface density 
of 800 undetected sources arcmin$^{-2}$ at an offset of 4.5\arcmin\ from 
\sgrastar, or a total of $2\times 10^{5}$ sources within 20 pc (9\arcmin) 
of \sgrastar.

There is no known class object of that could account for that 
large a number of hard X-ray sources at the Galactic center. In fact,
the total stellar mass within the inner 20 pc of the Galaxy is only 
$\sim 10^8$ \msun\ \citep{lzm02}, so in order to account for the diffuse 
emission, $\sim 0.2$\% of all stellar sources
would have to be hard X-ray sources with $L_{\rm X} > 3\times10^{29}$~\ergsec. 
The only plausible candidates for such a large population of hard X-ray sources
are cataclysmic variables (CVs) and young stellar objects 
(YSOs). 
We can estimate the number of CVs within 20 pc of the Galactic 
center by scaling their local density by the relative stellar mass density.
Based upon the mass model of \citet{lzm02}, the stellar density in 
the inner 20 pc is 1000 \msun\ pc$^{-3}$, compared to a local density 
of 0.1 \msun\ pc$^{-3}$. The local space density of CVs is 
at most $3 \times 10^{-5}$ pc$^{-3}$ \citep{war95,sch02}, so the space 
density at the Galactic center should be less than $0.03$ pc$^{-3}$. 
Therefore, we expect only $\sim 10^4$ CVs there, which is an order
of magnitude smaller than the number of point sources needed to explain 
the hard component of the diffuse emission. 

The number of YSOs would depend on the recent formation rate for low-mass 
stars. If stars have formed steadily over the $\sim 10$ Gyr lifetime of the 
Galaxy, then the $\sim 10^{8}$~\msun\ of stars within the inner 20~pc would
have formed at a rate of $\sim 10^{-2}$ y$^{-1}$, which would imply
that there should be $\sim 10^4$ stars younger than 1~Myr old. 
This number is comparable to the limit obtained from the fact that only 
$\sim 5$ hard X-ray sources in our field exhibit
flares with luminosities and durations that are similar to one seen from YSOs 
\citep{mun04}. About 0.1\% of YSOs exhibit flares that would have been
detectable from the Galactic center, which suggests that $\la 5000$
YSOs lie near the Galactic center \citep{fei04,gro04}, which is still far
too few to produce the hard diffuse emission.

Other X-ray sources that are similarly abundant are unlikely to contribute
to the hard component of the diffuse emission because their luminosities 
are too low or their spectra are too soft. 
For instance, although RS CVns are at least as 
numerous as CVs \citep{fms95}, they 
have thermal spectra with $kT < 2$ keV \citep{sdw96}.
Likewise, while isolated neutron stars 
accreting from the interstellar medium could make up $\sim 1$\% of the mass 
of the Galactic center \citep{zane96}, they must be less luminous 
than $10^{29}$~\ergsec, or else many thousands of similar systems would have 
been detected in the local Galaxy in the {\it ROSAT} All-Sky survey 
\citep{per03}. In contrast, less than a dozen candidate isolated neutron 
stars were identified with {\it ROSAT}.
Therefore, if the diffuse emission is produced by undetected point sources,
they would have to belong to a population of sources that has not yet
been identified.

\section{Conclusions}

Using a 600~ks exposure with the ACIS-I aboard \chandra, we have studied the 
spectrum of diffuse X-ray emission from several regions 
within a projected distance of 20 pc of \sgrastar.
The spectrum of the diffuse emission  exhibits He-like and 
H-like lines from Si, S, Ar, Ca, and Fe, as well as a prominent low-ionization
Fe line. If the spectrum is modeled as originating from diffuse plasma, 
two components with temperatures of 0.8~keV and 8~keV are required, along
with line emission from low-ionization Fe at 6.4~keV. The energies and 
flux ratios of the lines from both temperature components are consistent 
with emission from plasmas in collisional ionization equilibrium.

In Table~\ref{tab:summary}, we provide a summary of the origins of the 
X-ray emission between 2--8 keV from the inner 20 pc of the Galaxy.
By far the largest contribution to the luminosity 
of the Galactic center is from diffuse emission. 
In comparison, detected point sources contribute only 10\% to the luminosity
of the Galactic center, while discrete filamentary features contribute less 
than 5\% of the total luminosity of the inner 20~pc of the Galaxy. 
These results are potentially
useful for understanding the origin of diffuse X-ray emission from 
distant galaxies with quiescent central black holes. 
However, it is important to note that these observations of the Galactic 
center are strongly
affected by interstellar absorption with a column density of at least 
$6\times 10^{22}$ cm$^{-2}$. Therefore, the cool emission with $kT \la 0.5$ 
keV that produces most of the 0.5--8.0 keV flux from distant galaxies is 
obscured at the Galactic center. 
At the same time, \chandra\ observations of other 
galaxies are not sensitive to the $kT \approx 8$ keV plasma that dominates
the flux we observe from the Galactic center, because this hard emission
has a much lower surface brightness than the $kT \la 1$ keV emission where
the \chandra\ effective area is largest (0.5--3 keV), and it is difficult 
to resolve from bright X-ray binaries. Thus, these observations of the 
Galactic center provide a unique view of the hottest components of the ISM 
of galaxies.

The properties of the soft, $kT \approx 0.8$~keV plasma component of the 
diffuse emission vary significantly across the image, both in temperature 
between 0.7 and 0.9~keV, and in surface brightness 
between $2\times10^{-14}$ and $1.7\times10^{-13}$~\ergcmsarcmin.
The variation in the properties of the soft plasma suggest that it is 
relatively young, because differential rotation at the Galactic center
should destroy any coherent features within $< 10^6$ y. 
Supernovae probably supply most of 
this energy, although the winds from WR and early O stars could also
contribute.  Within the 
inner 20 pc of the Galaxy, the $\approx 3\times10^{36}$ erg s$^{-1}$ 
lost by the plasma through radiative cooling could be replaced by 
1\% of the kinetic energy of one supernova occurring every 
$\sim 10^5$ y. The inner 20 pc of our Galaxy contains about 0.1\%
of its total mass, so assuming that one supernova occurs every 100 y in the 
Galaxy, this rate is roughly consistent with that expected near the Galactic
center.

The hard component of the diffuse emission is more spatially uniform 
than the soft, but the intensities of the two components are still 
correlated. Although this might suggest a common origin for the two 
plasma components, supernovae and massive stars are not usually observed to
produce plasma with $kT \gtrsim 3$ keV. This hard 
emission is distributed throughout the Galactic plane, so
it is not likely to be associated with an outburst from \sgrastar.
Instead, the hard emission could result from a $kT\approx 8$~keV plasma 
that is heated indirectly by massive stars and
supernova remnants, which, for example, could drive reconnection in the 
magnetic fields near the Galactic center \citep{tan99}. However, a 8~keV 
thermal plasma would freely expand away from the Galactic center, and
would require $\approx 10^{40}$ erg s$^{-1}$ to 
sustain. This is equivalent to the entire kinetic energy of one supernova 
every 3000 years, which is a much larger rate than usually assumed for 
supernova. Supernova are the most energetic source of heat for the ISM, 
so if the hard diffuse emission is produced by a $kT \approx 8$ keV plasma,
it would imply that there is a significant shortcoming in our understanding 
of heating mechanisms for the ISM.

Alternative explanations for the hard diffuse emission that were intended to 
lessen the energy required are equally unsatisfying.
The suggestion that the hard diffuse emission 
originates from undetected stellar X-ray sources is unlikely because 
there is no known class of source that are numerous enough, bright enough,
and hot enough to produce the observed flux of $kT \approx 8$ keV diffuse 
emission. Likewise, if the hard diffuse emission originates from
non-thermal processes, such as the shocks that accelerate cosmic rays, 
the energies and ratios of the intensities of 
the line emission should deviate measurably from the values expected for
a plasma in thermal equilibrium \citep[e.g.,][]{mas02}. These deviations are 
not observed in our \chandra\ observations, which presents a challenge to the
current non-thermal models.

Further observations should clarify the nature
of the diffuse X-ray emission from the Galactic center. X-ray missions 
with higher spectral resolution, such as ASTRO-E 2, 
will be able to better-constrain the properties of the putative diffuse plasma
by resolving the individual transitions of He-like 
and H-like Fe, and possibly measuring the velocity dispersions of the Fe ions
themselves. Alternatively, a future hard X-ray survey, such as EXIST, could
identify a heretofore unknown population of numerous, faint, hard X-ray 
sources that may be responsible for producing the $kT \approx 8$ keV 
diffuse emission.

\acknowledgements{We are grateful to F. Paerls and T. Treu for helpful
discussions as this paper was taking shape, and the referee for insightful 
comments that aided us in clarifying our conclusions. MPM was supported by 
a Hubble 
Fellowship from the Space Telescope Science Institute, which is operated
by the Association of Universities for Research in Astronomy, Inc.,
under NASA contract NAS 5-26555.}

\begin{deluxetable}{lccccc}
\tablecolumns{6}
\tablewidth{0pc}
\tablecaption{Observations of the Inner 20 pc of the Galaxy\label{tab:obs}}
\tablehead{
\colhead{} & \colhead{} & \colhead{} & 
\multicolumn{2}{c}{Aim Point} & \colhead{} \\
\colhead{Start Time} & \colhead{Sequence} & \colhead{Exposure} & 
\colhead{RA} & \colhead{DEC} & \colhead{Roll} \\
\colhead{(UT)} & \colhead{} & \colhead{(s)} 
& \multicolumn{2}{c}{(degrees J2000)} & \colhead{(degrees)}
} 
\startdata
1999 Sep 21 02:43:00 & 0242  & 40,872 & 266.41382 & -29.0130 & 268 \\
2000 Oct 26 18:15:11 & 1561 & 35,705 & 266.41344 & -29.0128 & 265 \\
2001 Jul 14 01:51:10 & 1561 & 13,504 & 266.41344 & -29.0128 & 265 \\
2002 Feb 19 14:27:32 & 2951  & 12,370 & 266.41867 & -29.0033 & 91 \\
2002 Mar 23 12:25:04 & 2952  & 11,859 & 266.41897 & -29.0034 & 88 \\
2002 Apr 19 10:39:01 & 2953  & 11,632 & 266.41923 & -29.0034 & 85 \\
2002 May 07 09:25:07 & 2954  & 12,455 & 266.41938 & -29.0037 & 82 \\
2002 May 22 22:59:15 & 2943  & 34,651 & 266.41991 & -29.0041 & 76 \\
2002 May 24 11:50:13 & 3663  & 37,959 & 266.41993 & -29.0041 & 76 \\
2002 May 25 15:16:03 & 3392  & 166,690 & 266.41992 & -29.0041 & 76 \\
2002 May 28 05:34:44 & 3393  & 158,026 & 266.41992 & -29.0041 & 76 \\
2002 Jun 03 01:24:37 & 3665  & 89,928 & 266.41992 & -29.0041 & 76 
\enddata
\end{deluxetable}

\begin{deluxetable}{llccc}
\tablecolumns{5}
\tablewidth{0pc}
\tablecaption{Extraction Regions for the Diffuse Emission\label{tab:regions}}
\tablehead{
\colhead{Region} & \colhead{Area} & \colhead{Offset} & \colhead{Net Counts} &
\colhead{Background} \\
\colhead{} & \colhead{(arcmin$^2$)} & \colhead{(arcmin)} & \colhead{} & 
\colhead{}
}
\startdata
Southeast & 34.7 & 7.5 & $1.44 \times 10^5$ & $0.42 \times 10^5$ \\
Southwest & 14.0 & 7.7 & $0.56 \times 10^5$ & $0.17 \times 10^5$ \\
Northwest & 23.9 & 7.8 & $0.82 \times 10^5$ & $0.27 \times 10^5$ \\
East & 12.0 & 7.8 & $0.65 \times 10^5$ & $0.16 \times 10^5$ \\
Close & 9.0 & 3.9 & $0.61 \times 10^5$ & $0.12 \times 10^5$ \\ [5pt]
Northeast & 46.0 & 7.6 & $4.74 \times 10^5$ & $0.53 \times 10^5$
\enddata
\end{deluxetable}

\begin{deluxetable}{lcccccc}
\tabletypesize{\scriptsize}
\tablecolumns{7}
\tablewidth{0pc}
\tablecaption{Two-$kT$ Plasma Model of Diffuse Emission\label{tab:twokt}} 
\tablehead{
\colhead{} & \colhead{Southeast} & \colhead{Southwest} & \colhead{Northwest} &
\colhead{East} & \colhead{Close} & \colhead{Northeast}\\
}
\startdata
$N_{\rm H,s}$ ($10^{22}$ cm$^{-2}$) &  1.4$_{- 0.1}^{+ 0.1}$ &  1.0$_{- 0.2}^{+ 0.1}$ &  1.7$_{- 0.2}^{+ 0.2}$ &  2.2$_{- 0.2}^{+ 0.1}$ &  1.6$_{- 0.2}^{+ 0.1}$ &  7.2$_{- 0.1}^{+ 0.1}$  \\
$N_{\rm H,pc,s}$ ($10^{22}$ cm$^{-2}$) &  4.8$_{- 0.3}^{+ 0.3}$ &  4.4$_{- 1.0}^{+ 1.0}$ &  4.9$_{- 0.5}^{+ 0.6}$ &  4.7$_{- 0.4}^{+ 0.4}$ &  4.8$_{- 0.6}^{+ 0.7}$ &  0.1$_{- 0.1}^{+11.2}$  \\
$f_{\rm s}$ (fixed) & 0.95 & 0.95 & 0.95 & 0.95 & 0.95 & 0.94  \\
$kT_{\rm s}$ (keV) &  0.82$_{-0.03}^{+0.03}$ &  0.71$_{-0.05}^{+0.05}$ &  0.70$_{-0.03}^{+0.03}$ &  0.92$_{-0.03}^{+0.08}$ &  0.80$_{-0.03}^{+0.07}$ &  0.88$_{-0.00}^{+0.02}$  \\
$K_{\rm EM,s}$ ($10^{-4}$ cm$^{-6}$ pc) &   2.1$_{- 0.3}^{+ 0.2}$ &   1.3$_{- 0.4}^{+ 0.3}$ &   2.2$_{- 0.5}^{+ 0.6}$ &   5.9$_{- 1.4}^{+ 1.2}$ &   3.0$_{- 0.8}^{+ 0.8}$ &  22.0$_{- 0.9}^{+ 1.1}$  \\
$F_{\rm s}$ ($10^{-13}$ erg cm$^{-2}$ s$^{-1}$ arcmin$^{-2}$) &  0.3 &  0.2 &  0.2 &  0.6 &  0.4 &  1.7  \\
$uF_{\rm s}$ ($10^{-13}$ erg cm$^{-2}$ s$^{-1}$ arcmin$^{-2}$) &  0.8 &  0.4 &  0.7 &  2.1 &  1.1 &  5.6  \\
$N_{\rm H,h}$ ($10^{22}$ cm$^{-2}$) & 4.8$_{-0.2}^{+0.3}$ & 4.0$_{-0.3}^{+0.5}$ & 6.1$_{-0.4}^{+0.4}$ & 4.7$_{-0.3}^{+0.6}$ & 4.7$_{-0.3}^{+0.4}$ & 1.16$_{-0.04}^{+0.03}$  \\
$N_{\rm H,pc,h}$ ($10^{22}$ cm$^{-2}$) &  83$_{- 20}^{+ 25}$ &  29$_{-  6}^{+ 13}$ &  72$_{- 15}^{+ 24}$ &  70$_{- 13}^{+ 40}$ &  28$_{-  6}^{+  9}$ &  14.7$_{-  0.4}^{+  1.3}$  \\
$f_{\rm h}$ & 0.50$_{0.17}^{+0.14}$ & 0.52$_{0.05}^{+0.05}$ & 0.60$_{0.09}^{+0.11}$ & 0.58$_{0.13}^{+0.19}$ & 0.48$_{0.05}^{+0.05}$ & 0.95$_{0.03}^{+0.04}$  \\
$kT_{\rm h}$ (keV) &   7.1$_{- 0.8}^{+ 0.2}$ &   8.1$_{- 0.6}^{+ 0.6}$ &   8.8$_{- 0.5}^{+ 0.7}$ &   5.9$_{- 1.1}^{+ 0.9}$ &   8.6$_{- 0.4}^{+ 0.7}$ &   7.9$_{- 0.2}^{+ 0.3}$  \\
$K_{\rm EM,h}$ ($10^{-4}$ cm$^{-6}$ pc) &    1.8$_{-  0.5}^{+  0.6}$ &    1.5$_{-  0.2}^{+  0.3}$ &    2.2$_{-  0.4}^{+  0.8}$ &    2.5$_{-  0.7}^{+  1.8}$ &    2.2$_{-  0.2}^{+  0.3}$ &    2.7$_{-  0.1}^{+  0.0}$  \\
$F_{\rm h}$ ($10^{-13}$ erg cm$^{-2}$ s$^{-1}$ arcmin$^{-2}$) &   1.5 &   1.6 &   1.5 &   1.6 &   2.5 &   2.4  \\
$uF_{\rm h}$ ($10^{-13}$ erg cm$^{-2}$ s$^{-1}$ arcmin$^{-2}$) &   4.0 &   3.5 &   5.1 &   5.9 &   5.3 &   5.6  \\
$Z_{\rm Si}/Z_{{\rm Si},\odot}$ & 1.0$_{-0.1}^{+0.1}$ & 1.5$_{-0.3}^{+0.5}$ & 1.5$_{-0.2}^{+0.3}$ & 0.7$_{-0.1}^{+0.2}$ & 1.3$_{-0.3}^{+0.3}$ & 0.80$_{-0.04}^{+0.03}$  \\
$Z_{\rm S}/Z_{{\rm S},\odot}$ & 2.5$_{-0.2}^{+0.4}$ & 2.7$_{-0.6}^{+0.8}$ & 2.2$_{-0.3}^{+0.5}$ & 1.6$_{-0.3}^{+0.3}$ & 2.3$_{-0.5}^{+0.6}$ & 1.03$_{-0.01}^{+0.04}$  \\
$Z_{\rm Ar}/Z_{{\rm Ar},\odot}$ & 3.6$_{-0.4}^{+0.6}$ & 4.1$_{-1.1}^{+1.2}$ & 2.8$_{-0.6}^{+0.8}$ & 1.8$_{-0.4}^{+0.4}$ & 3.3$_{-0.8}^{+0.9}$ & 1.08$_{-0.05}^{+0.06}$  \\
$Z_{\rm Ca}/Z_{{\rm Ca},\odot}$ & 2.4$_{-0.5}^{+0.5}$ & 2.3$_{-1.1}^{+1.3}$ & 1.3$_{-0.8}^{+0.8}$ & 2.0$_{-0.5}^{+0.4}$ & 2.4$_{-0.8}^{+0.8}$ & 1.1$_{-0.1}^{+0.1}$  \\
$Z_{\rm Fe}/Z_{{\rm Fe},\odot}$ & 0.7$_{-0.2}^{+0.1}$ & 0.8$_{-0.1}^{+0.1}$ & 0.7$_{-0.1}^{+0.1}$ & 0.7$_{-0.2}^{+0.2}$ & 0.8$_{-0.1}^{+0.1}$ & 0.57$_{-0.02}^{+0.04}$  \\
Fe K-$\alpha$ ($10^{-7}$ ph cm$^{-2}$ s$^{-1}$ arcmin$^{-2}$) &  11$_{- 2}^{+ 4}$ &   9$_{- 1}^{+ 1}$ &  15$_{- 2}^{+ 5}$ &  14$_{- 3}^{+ 9}$ &  14$_{- 1}^{+ 2}$ &  42$_{- 1}^{+ 1}$  \\
$\chi^2/\nu$ &  494.0/462 &  484.4/462 &  519.1/462 &  621.8/462 &  428.5/462 &  519.9/462
\enddata
\tablecomments{The fluxes $F$ and $uF$ are between 2--8~keV. The former is 
the observed flux, the latter is the de-absorbed flux.}
\end{deluxetable}

\begin{deluxetable}{lcccccc}
\tabletypesize{\scriptsize}
\tablecolumns{7}
\tablewidth{0pc}
\tablecaption{Properties of Iron Emission\label{tab:iron}} 
\tablehead{
\colhead{} & \colhead{Southeast} & \colhead{Southwest} & \colhead{Northwest} &
\colhead{East} & \colhead{Close} & \colhead{Northeast}\\
}
\startdata
$\Gamma$   &  1.1$_{-0.1}^{+0.1}$ &  0.5$_{-0.2}^{+0.2}$ &  0.5$_{-0.2}^{+0.1}$ &  1.1$_{-0.2}^{+0.2}$ &  0.8$_{-0.2}^{+0.1}$ &  1.4$_{-0.1}^{+0.0}$ \\
$N_{\Gamma}$  ($10^{-5}$ ph cm$^{-2}$ s$^{-1}$ keV$^{-1}$ arcmin$^{-2}$) &  1.9$_{- 0.6}^{+ 0.4}$ &  0.7$_{- 0.2}^{+ 0.3}$ &  0.7$_{- 0.2}^{+ 0.2}$ &  2.2$_{- 0.7}^{+ 0.9}$ &  2.1$_{- 0.5}^{+ 0.6}$ &  7.3$_{- 0.6}^{+ 0.6}$ \\
$E_{{\rm Fe K-}\alpha}$  (keV) & 6.39$_{-0.01}^{+0.01}$ & 6.37$_{-0.03}^{+0.03}$ & 6.443$_{-0.008}^{+0.011}$ & 6.38$_{-0.04}^{+0.04}$ & 6.40$_{-0.02}^{+0.01}$ & 6.395$_{-0.002}^{+0.002}$ \\
$W_{{\rm Fe K-}\alpha}$  (eV)\tablenotemark{a} &  $<70$ &  40&  40&  40&  40&  37$^{+9}_{-10}$\\
$I_{{\rm Fe K-}\alpha}$  (10$^{-7}$ ph cm$^{-2}$ s$^{-1}$ arcmin$^{-2}$) &   4.0$_{- 0.4}^{+ 0.5}$ &   3.9$_{- 0.8}^{+ 0.8}$ &   7.6$_{- 0.5}^{+ 0.9}$ &   5.9$_{- 1.3}^{+ 1.0}$ &   8.4$_{- 0.8}^{+ 1.0}$ &  30.9$_{- 0.5}^{+ 0.5}$ \\
$E_{{\rm Fe He-}\alpha}$  (keV) & 6.637$_{-0.008}^{+0.006}$ & 6.665$_{-0.020}^{+0.009}$ & 6.73$_{-0.02}^{+0.01}$ & 6.670$_{-0.02}^{+0.02}$ & 6.658$_{-0.017}^{+0.007}$ & 6.668$_{-0.005}^{+0.004}$ \\
$W_{{\rm Fe He-}\alpha}$  (eV)\tablenotemark{a} &  $<40$ &  40&  40&  40&  40&  $50^{+15}_{-30}$\\
$I_{{\rm Fe He-}\alpha}$  (10$^{-7}$ ph cm$^{-2}$ s$^{-1}$ arcmin$^{-2}$) &  12.2$_{- 0.7}^{+ 0.5}$ &  12$_{- 1}^{+ 1}$ &  10.3$_{- 1.5}^{+ 0.7}$ &  16$_{- 2}^{+ 1}$ &  17$_{- 2}^{+ 1}$ &  16.8$_{- 0.6}^{+ 0.6}$ \\
$E_{{\rm Fe H-}\alpha}$  (keV) & 6.86$_{-0.05}^{+0.05}$ & 6.98$_{-0.08}^{+0.03}$ & 6.94$_{-0.04}^{+0.03}$ & 6.97\tablenotemark{b} & 6.94$_{-0.04}^{+0.01}$ & 6.948$_{-0.007}^{+0.006}$ \\
$W_{{\rm Fe H-}\alpha}$  (eV) &   0&   0&   0&   0&   0&   0\\
$I_{{\rm Fe H-}\alpha}$  (10$^{-7}$ ph cm$^{-2}$ s$^{-1}$ arcmin$^{-2}$) &   2.8$_{- 0.4}^{+ 1.0}$ &   4.7$_{- 1.4}^{+ 1.2}$ &   4.2$_{- 1.3}^{+ 0.5}$ & $< 2.7$ &   8.2$_{- 1.5}^{+ 1.5}$ &   6.7$_{- 0.7}^{+ 0.7}$ \\
$I_{{\rm K-}\alpha}/I_{{\rm He-}\alpha}$   &  0.3$_{- 0.0}^{+ 0.1}$ &  0.3$_{- 0.1}^{+ 0.1}$ &  0.7$_{- 0.1}^{+ 0.2}$ &  0.4$_{- 0.1}^{+ 0.1}$ &  0.5$_{- 0.1}^{+ 0.1}$ &  1.8$_{- 0.1}^{+ 0.1}$ \\
$I_{{\rm He-}\alpha}/I_{{\rm H-}\alpha}$   &  4.4$_{- 1.6}^{+ 1.2}$ &  2.5$_{- 0.8}^{+ 1.8}$ &  2.5$_{- 0.5}^{+ 1.4}$ & $> 4.3$ &  2.1$_{- 0.5}^{+ 0.7}$ &  2.5$_{- 0.3}^{+ 0.5}$ \\
$\chi^2/\nu$   &  204.1/230 &  245.6/230 &  236.8/230 &  315.0/231 &  219.9/230 &  240.6/230
\enddata   
\tablenotetext{a}{The widths of the K-$\alpha$ and He-$\alpha$ lines were only
allowed to vary for the spectra from the Northeast and Southeast.}
\tablenotetext{b}{The H-$\alpha$ line was not detected in the East, so its
centroid was fixed in order to obtain an upper limit to the line flux.}
\end{deluxetable}

\begin{deluxetable}{lcccccc}
\tablecolumns{7}
\tablewidth{0pc}
\tablecaption{Flux from the Plasma Components and Point Sources\label{tab:difflux}} 
\tablehead{
\colhead{} & \colhead{Southeast} & \colhead{Southwest} & \colhead{Northwest} &
\colhead{East} & \colhead{Close} & \colhead{Northeast}\\
\multicolumn{7}{c}{Total Fluxes from Regions}
}
\startdata
$F_{\rm tot}$ &  1.8 &  1.8 &  1.7 &  2.3 &  3.0 &  4.5  \\
$F_{\rm detected}$ & 0.2 & 0.3 & 0.2 & 0.2 & 1.0 &  0.4 \\
\cutinhead{Component Fluxes Assuming No Undetected Point Source Component}
$F_{\rm s}$ &  0.3 &  0.2 &  0.2 &  0.6 &  0.4 &  1.8  \\
$F_{\rm h}$ &  1.5 &  1.6 &  1.5 &  1.6 &  2.5 &  2.6  \\
\cutinhead{Component Fluxes with Undetected Point Source Component}
$F_{\rm s}$ &  0.5 &  0.4 &  0.2 &  0.5 &  0.5 &  1.7  \\
$F_{\rm h}$ &  0.0 &  0.0 &  0.8 &  0.9 &  0.1 &  1.9  \\
$F_{\rm undetected}$ &  1.3 &  1.4 &  0.7 &  0.8 &  2.4 &  0.9 
\enddata
\tablecomments{$F_{\rm tot}$ denotes the total diffuse flux, 
$F_{\rm detected}$ the total flux from detected point sources, 
$F_{\rm s}$ the soft component of the diffuse flux, $F_{\rm h}$ the 
hard component, and $F_{\rm undetected}$ the maximum contribution from 
undetected point sources. 
All fluxes are those observed between 2--8~keV, in units of 
$10^{-13}$~\ergcmsarcmin.}
\end{deluxetable} 

\begin{deluxetable}{lccccccc}
\tabletypesize{\scriptsize}
\tablecolumns{8}
\tablewidth{0pc}
\tablecaption{Properties of the Plasma Components\label{tab:prop}} 
\tablehead{
\colhead{} & \colhead{Scaling} & 
\colhead{Southeast} & \colhead{Southwest} & \colhead{Northwest} &
\colhead{East} & \colhead{Close} & \colhead{Northeast}}
\startdata
\cutinhead{Soft Component}
$L_{\rm X}$ ($10^{33}$ erg s$^{-1}$ arcmin$^{-2}$) &  &  11.8 &   6.3 &  10.1 &  32.8 &  16.4 &  87.1  \\
$kT$ (keV) &  &  0.8 &  0.7 &  0.7 &  0.9 &  0.8 &  0.9  \\
$n$ (cm$^{-3}$) & $d_{50}^{-1/2}$ & 0.1 & 0.1 & 0.2 & 0.2 & 0.2 & 0.5  \\
$M$ ($M_{\odot}$ arcmin$^{-2}$) & $d_{50}^{1/2}$ &  1.0 &  0.8 &  1.0 &  1.6 &  1.2 &  3.1  \\
$U$ ($10^{-9}$ erg cm$^{-3}$) & $d_{50}^{-1/2}$ &  0.3 &  0.2 &  0.3 &  0.5 &  0.3 &  1.0  \\
$E$ ($10^{48}$ erg arcmin$^{-2}$) & $d_{50}^{1/2}$ &  2 &  2 &  2 &  4 &  3 &  8  \\
$t_{\rm cool}$ ($10^7$ yr) & $d_{50}^{3/2}$ &  0.6 &  0.8 &  0.6 &  0.4 &  0.5 &  0.3  \\
$c_s$ (km s$^{-1}$) &  &   510 &   470 &   470 &   540 &   500 &   530  \\
$t_{\rm exp}$ ($10^5$ yr) & $d_{50}$ &  0.9 &  1.0 &  1.0 &  0.9 &  1.0 &  0.9  \\
$\dot{W}$ ($10^{36}$ erg s$^{-1}$) & $d_{50}^{1/6}$ &     4 &     2 &     3 &     8 &     4 &    14  \\
$t_{PdV}$ ($10^{4}$ yr) & $d_{50}^{1/3}$ &    2 &    2 &    2 &    2 &    2 &    2  \\
\cutinhead{Hard Component}
$L_{\rm X}$ ($10^{33}$ erg s$^{-1}$ arcmin$^{-2}$) &  &   6.2 &   5.4 &   7.9 &   9.1 &   8.2 &   8.7  \\
$kT$ (keV) &  &  7.1 &  8.1 &  8.8 &  5.9 &  8.6 &  7.9  \\
$n$ (cm$^{-3}$) & $d_{50}^{-1/2}$ & 0.1 & 0.1 & 0.2 & 0.2 & 0.2 & 0.2  \\
$M$ ($M_{\odot}$ arcmin$^{-2}$) & $d_{50}^{1/2}$ &  0.9 &  0.8 &  1.0 &  1.1 &  1.0 &  1.1  \\
$U$ ($10^{-9}$ erg cm$^{-3}$) & $d_{50}^{-1/2}$ &  2 &  2 &  3 &  2 &  3 &  3  \\
$E$ ($10^{48}$ erg arcmin$^{-2}$) & $d_{50}^{1/2}$ & 18 & 19 & 24 & 18 & 24 & 25  \\
$t_{\rm cool}$ ($10^7$ yr) & $d_{50}^{3/2}$ &  10 & 11 & 10 &  6 &  10 &  9  \\
$c_s$ (km s$^{-1}$) &  &  1500 &  1600 &  1700 &  1400 &  1600 &  1600  \\
$t_{\rm exp}$ ($10^5$ yr) & $d_{50}$  &  0.3 &  0.3 &  0.3 &  0.4 &  0.3 &  0.3  \\
$\dot{W}$ ($10^{36}$ erg s$^{-1}$) & $d_{50}^{1/6}$ &    90 &   100 &   136 &    80 &   134 &   130  \\
$t_{PdV}$ ($10^{4}$ yr) & $d_{50}^{1/3}$ &    0.6 &    0.6 &    0.6 &    0.7 &    0.6 &    0.6  \\
\enddata
\end{deluxetable}

\begin{deluxetable}{lccc}
\tablecolumns{4}
\tablewidth{0pc}
\tablecaption{Summary of X-rays from the Galactic Center\label{tab:summary}} 
\tablehead{
\colhead{Source} & \colhead{Size} & \colhead{$L_{\rm X}$} & 
\colhead{References}\\
\colhead{} & \colhead{(arcmin$^2$)} & \colhead{(\ergsec; 2--8 keV)} & \colhead{}
}
\startdata
\sgrastar\ & $7\times10^{-5}$ & $2\times 10^{33}$ & [1] \\
Central Stellar Cluster & 0.08 & $2\times10^{33}$ & [1]\\
Sgr A East & 1.4 & $8\times10^{34}$ & [2] \\
Bipolar Outflow & 30 & $2\times10^{34}$ & [3] \\ 
Non-thermal Filaments & 0.04 & $1\times 10^{34}$ & [4] \\
Neutral Iron Filaments & 0.5 & $2\times 10^{34}$ & [5] \\
Detected Point Sources & 290 & $1\times10^{35}$ & [6] \\ 
Diffuse Emission & 290 &  $2\times10^{36}$ & [7] \\
\enddata
\tablerefs{[1] \citet{bag03}; [2] \citet{mae02}; [3] \citet{mor03} 
[4] F. Yusef-Zadeh \etal, 
in preparation; [5] \citet{par03}; [6] \citet{mun03}; [7] this work}
\end{deluxetable}

\end{document}